\documentclass[journal]{IEEEtran}

\usepackage{amssymb}
\usepackage{amsthm}
\usepackage{amsmath}
\usepackage{xcolor}
\usepackage{comment}
\usepackage{nomencl}
\usepackage{ifthen}
\usepackage{etoolbox}
\usepackage{bbm}
\usepackage{algorithm}
\usepackage{algpseudocode}
\usepackage{enumerate}
\usepackage{float}
\usepackage{booktabs}
\usepackage{tikz}
\usepackage{subcaption}
\usepackage{soul}
\usepackage{multirow}
\usepackage{algpseudocode}
\usepackage{graphicx}
\usepackage{epstopdf}
\usepackage{comment}
\usepackage{titlesec}
\usepackage{url}

\definecolor{JHUblue}{RGB}{0, 45, 114}
\definecolor{JHUwhite}{RGB}{255, 255, 255}  
\definecolor{JHUblack}{RGB}{ 0, 0, 0}  
\definecolor{color1}{RGB}{215, 38, 56}
\definecolor{color2}{RGB}{63, 136, 197}
\definecolor{color4}{RGB}{62, 58, 83}
\definecolor{plot1}{RGB}{31, 119, 180}
\definecolor{plot2}{RGB}{255, 127, 14}
\definecolor{plot3}{RGB}{44, 160, 44}
\definecolor{amber}{rgb}{1.0, 0.49, 0.0}

\algtext*{EndIf}
\algtext*{EndFor}
\titlespacing*{\section}{0pt}{1.0ex plus 0.2ex minus 0.2ex}{0.6ex}
\titlespacing*{\subsection}{0pt}{0.8ex plus 0.2ex minus 0.2ex}{0.4ex}
\titlespacing*{\subsubsection}{0pt}{0.6ex plus 0.2ex minus 0.2ex}{0.3ex}

\allowdisplaybreaks

\makenomenclature
\renewcommand{\nomgroup}[1]{%
  \item[%
    \ifthenelse{\equal{#1}{A}}{A. \textit{Sets}}{}%
    \ifthenelse{\equal{#1}{B}}{B. \textit{Parameters}}{}%
    \ifthenelse{\equal{#1}{C}}{C. \textit{Variables}}{}%
    ]%
    \vspace{10pt}\hspace*{-\leftmargin}\vspace{10pt}%
}

\newtheorem{theorem}{Theorem}

\theoremstyle{plain}

\allowdisplaybreaks 
\begin{document}
\title{Reachability Guarantees for Energy Arbitrage}
%
\author{Tomás Tapia and 
        Yury Dvorkin
\vspace{-10mm}
}

 
\maketitle

\nomenclature[A]{$\mathcal{T}$}{Set of time periods $\{1,...,T\}$, indexed by $t$.}%
\nomenclature[A]{$\mathcal{T}^{\text{target}}$}{Subset of times at which the chance constraint must hold.}%

\nomenclature[C]{$e_t$}{State-of-charge or energy stored at time $t$.}
\nomenclature[C]{$c_t$}{Energy charged into the battery at time $t$.}
\nomenclature[C]{$d_t$}{Energy discharged from the battery at time $t$.}
\nomenclature[C]{$a_t$}{Battery control action at time $t$, that takes value $1$ if charge, $-1$ if discharge, or $0$ if idle.}
\nomenclature[C]{$\tau^{\star}$}{Minimum stopping time when the battery starts to preserve energy.}

\nomenclature[B]{$\eta$}{Efficiency factor [\%].}
\nomenclature[B]{$E_{\min}$}{Battery minimum capacity [MWh].}
\nomenclature[B]{$E_{\max}$}{Battery maximum capacity [MWh].}
\nomenclature[B]{$P$}{Maximum charge and discharge rate [MW].}
\nomenclature[B]{$\boldsymbol{\lambda}_t$}{Uncertain energy price at time $t$ [USD/MWh].}
\nomenclature[B]{$e^{\text{target}}$}{Energy target at the end-of-horizon time $T$ [MWh].}
\nomenclature[B]{$\epsilon$}{Risk tolerance for the chance constraint [\%].}
\nomenclature[B]{$\delta$}{Band tolerance for the energy target [MWh].}

\begin{abstract}
Batteries performing energy arbitrage may overdischarge ahead of a high-demand or critical operating hour when they re-optimize in real time and depart from the day-ahead schedule. This paper introduces a framework for battery energy arbitrage under uncertain market prices that integrates chance-constrained terminal state-of-charge (SoC) requirements with online threshold policies. We first cast the multi-interval arbitrage problem as a stochastic dynamic program enhanced by a probabilistic end-of-horizon SoC constraint, ensuring with high confidence that the battery terminates within a prescribed energy band. We then apply a $k$-search algorithm to derive explicit charging (buying) and discharging (selling) thresholds with a provable worst-case competitive ratio, and compute the corresponding action probabilities over the decision horizon. To compute exact distributions under operational limits, we develop a probability redistribution pruning method and use it to quantify the likelihood of meeting the terminal SoC band. Leveraging the resulting SoC distribution, we estimate the minimum stopping time required to satisfy the SoC chance constraint. Computational experiments on historical real price data demonstrate that the proposed framework improves SoC estimation and supports chance-constraint satisfaction, while also showing that reachability is highly sensitive to a given initial SoC and operating start time.
\end{abstract}


\IEEEpeerreviewmaketitle

\vspace{-2mm}
\section{Introduction} \label{Sec1:Introduction}

\IEEEPARstart{M}{ulti}-period optimization is indispensable for the efficient operation of energy-constrained resources, such as battery storage, which are characterized by complex intertemporal state-of-charge (SoC) dynamics and operational limits. For both asset- and system-level perspectives, decision-making processes must rigorously account for economic trade-offs, which are inherently influenced by operational variability, uncertainty, and risk tolerance, as well as for the need to reconcile asset- and system-level objectives. 

From a battery perspective, energy arbitrage requires sequential charge and discharge decisions based on current and forecast electricity prices. This decision problem may lead to multiple admissible SoC trajectories that reach the same terminal SoC. However, these trajectories may differ in arbitrage revenue and in when the battery has energy available to the system. From a system perspective, therefore, what matters is not only the terminal SoC value but also the trajectory taken to it. The difference between admissible trajectories becomes relevant because forecast errors in renewable generation, load, and prices lead batteries to revise their decisions in real time, a step we refer to below as \emph{replanning}.

The main operational concern is that economically attractive real time actions in the short term may reduce the battery's ability to respond later. In particular, real time (RT) price spikes may lead the battery to overdischarge before \emph{critical hours}, leaving insufficient SoC when the system experiences tight supply, high demand, or elevated reliability risk. Conversely, excessive charging may reduce operational flexibility and increase congestion risk, while premature inactivity may leave insufficient flexibility to respond to subsequent conditions. Although battery storage is technically well suited to manage such variability, the day-ahead (DA) and RT market software tools typically rely on deterministic forecasts without explicit mechanisms to hedge against real time volatility \cite{bienstock2024risk}. In this paper, we therefore link sequential arbitrage decisions to the probability of reaching a prescribed terminal SoC target.

System-wide storage adoption and critical-hour risks are also related. As grid-scale battery deployment increases, the system relies more heavily on storage during critical hours. A battery that overdischarges beforehand may not be able to respond as expected by the system operator, forcing costly ad hoc out-of-market interventions rather than a market-based response. System operators have responded to this risk by managing SoC requirements administratively. The California Independent System Operator (CAISO) introduced a Minimum State‐of‐Charge (MSOC) program in 2022. This mechanism was not integrated into the market-clearing routine and functioned as an administrative overlay on the DA market, requiring certain batteries to reserve a fraction of their energy to remain available during peak-demand and evening ramp periods \cite{CAISO2024}. Although MSOC expired in 2023 and was not extended due to operational improvements in forecast processes and bidding behavior, this rule was viewed as a short-term measure to maintain grid reliability with some economic and efficiency limitations \cite{CAISO_DMM_MSOC_Comments_2023}. However, its underlying objective can still be replicated through improved exceptional dispatch, which compensates batteries for missed market opportunities and incentivizes energy retention during high-value intervals. Similarly, in 2022, the Electric Reliability Council of Texas (ERCOT) approved the Nodal Protocol Revision Request 1186 (NPRR1186), which introduced requirements to track and monitor SoC, to ensure compliance with ancillary service obligations, and to maintain a minimum SoC so that system operators can estimate deliverability for the whole committed period \cite{PUCT2024}. However, MSOC and NPRR1186 are administrative mechanisms that operate outside the market-clearing process and are thus prone to the limitations of ad hoc out-of-market interventions \cite{al2014role}. This is precisely the gap our chance-constrained alternative is designed to address by embedding the probabilistic terminal SoC requirement in the battery's replanning decisions.

We propose a framework to reconcile the system and asset perspectives by quantifying the probability of remaining within an admissible terminal SoC band. This framework employs a competitive threshold policy motivated by policy requirements aimed at ensuring the deliverability of energy arbitrage, regulation and reserve services. This approach enables us to determine the battery's \textit{stopping time}, the point at which it must start preserving energy to avoid violating terminal SoC constraints, and the associated cost of stopping. This framework provides system operators with sufficient advance notice to take corrective actions and re-dispatch the system.

\begin{figure}[ht]
    \centering
    \includegraphics[width=1.07\linewidth, trim={0mm 0mm 5mm 0mm},clip]{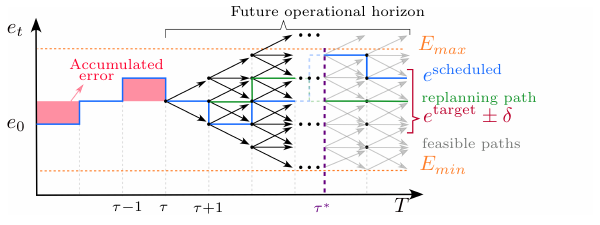}
    \caption{SoC trajectories for the battery arbitrage problem.}
    \label{fig:enter-label2}
\end{figure}

We illustrate the main idea behind the proposed framework using Fig. \ref{fig:enter-label2}, which displays the evolution of the SoC ($e_t$) for a given battery storage asset through a finite time horizon $T$. Value $e_{0}$ corresponds to the initial SoC, and value $e^{\rm target}$ is the SoC target band (e.g., as determined by the DA schedule or by other means), while $\delta$ is a user-defined tolerance margin (e.g., set by the system policy). Together, the red bracket at $T$ highlights the required terminal SoC band $\mathcal{E} = [e^{\rm target} \pm \delta]$. The solid blue line represents the realized SoC trajectory $e_{t}$ up to some \emph{replanning} time $\tau$, with red blocks indicating the accumulated error from the original schedule. Before $\tau^{\star}$, the battery follows this realized (unrestricted) trajectory and all feasible charging and discharging actions are available at every time-step. After $\tau$, the blue line represents the DA-planned SoC trajectory with the aim of reaching the terminal SoC band. At time $\tau$, each decision \(a_{t}\in\{-1,0,+1\}\) (discharge, idle, or charge) generates three branches (black arrows) in a decision tree, capturing all possible future trajectories. Gray branches denote trajectories that are still feasible but that may also be subsequently pruned because they violate the SoC capacity limits or fail to reach $\mathcal{E}$ by $T$.
Finally, the orange dotted lines indicate the minimum ($E_{\rm min}$) and maximum ($E_{\rm max}$) SoC capacity limits, respectively. Thus, we develop an approach that finds time $\tau^{\star}$ and the replanning trajectory (green arrows in Fig. \ref{fig:enter-label2}) that assures the required terminal SoC band ($\mathcal{E}$) with a given probability. From $\tau^{\star}$ onward, the battery is restricted to the subset of actions consistent with the green (surviving) branches, i.e., it switches from an unrestricted policy to a constrained energy-preservation mode. To this end, we develop an approach to optimize the stopping time criterion to begin preserving energy to meet a given terminal SoC band target. We also provide an explicit performance guarantee that quantifies the probability that the battery will remain within a target SoC band under uncertain price trajectories.

\subsection{Literature review}
In the past decade, optimization techniques for battery management have advanced rapidly \cite{ahmad2025renewable}. The authors in \cite{sioshansi2021energy} present a broad survey of such techniques and outline research challenges. In the context of \cite{sioshansi2021energy}, four research areas are directly relevant to our framework: (i) valuation of stored energy, (ii) profitability and market-operation studies, (iii) online threshold policies, and (iv) uncertainty quantification for storage.

\emph{(i) Valuation of stored energy.} One of the major current challenges is estimating the value of stored energy for arbitrage applications. In this context, \cite{sioshansi2009estimating} estimates the historical annual value of stored energy for a price-taking storage system performing arbitrage in the PJM market, considering factors such as fuel prices, transmission line constraints, battery efficiency, and capacity limits. Similarly, \cite{mcconnell2015estimating} estimates the value of stored energy in Australia's National Electricity Market considering arbitrage opportunities and capacity service provision. The authors in \cite{xu2020operational} present an analytical formulation for the value of energy storage for a price-taker battery asset under multistage uncertainty. These studies value a given SoC trajectory ex-post, but do not provide an operational decision rule that guarantees reaching a prescribed terminal SoC.

\emph{(ii) Profitability and market-operation studies.} Beyond the valuation of stored energy, recent studies have focused on analyzing the profitability of energy storage through different market structures. For instance, \cite{dvorkin2016ensuring} analyzes the profitability of battery storage investments, including an arbitrage analysis to determine the optimal location and sizing of energy storage systems. The authors in \cite{zhang2021arbitrage} propose a strategic framework for energy arbitrage and evaluate the performance of different energy storage technologies, accounting for variations in overall efficiency and life-cycle costs, and \cite{singhal2020pricing} introduces alternative SoC market models and examines arbitrage opportunities in different price settings. These studies characterize storage's performance within a given market structure, but evaluate a specified policy rather than deriving one with a worst-case or chance-constrained terminal-state guarantee.

The authors in \cite{krishnamurthy2017energy} examine energy arbitrage opportunities for grid-level battery storage in the DA market, allowing corrective actions in the RT market. Similarly, \cite{zheng2022arbitraging} proposes an analytical stochastic dynamic programming (SDP) algorithm to optimize variable-efficiency energy storage for price arbitrage in the RT market. The authors in \cite{chen2022battery} analyze the integration of battery storage in DA security-constrained unit-commitment, while \cite{qi2023chance, qi2024chance} present chance-constrained pricing models for energy storage systems. Furthermore, the authors in \cite{zhang2023day} introduce a cooperative DA dispatch framework in which energy storage provides multiple regulation services to improve system reliability, and \cite{cheng2016co} introduces a nested Markov decision process model that co-optimizes energy storage for joint energy arbitrage and frequency regulation. More closely related to our terminal-SoC reachability focus, \cite{baldick2026} proposes a market design in which storage assets bid only on their end-of-horizon SoC deviation.

\emph{(iii) Online threshold policies.} To cope with the impacts of forecast errors, a rich literature on online policies has emerged. Its foundation lies in classical threshold policies, such as the secretary problem \cite{babaioff2008online,degroot1968some} and stopping time theory \cite{degroot1968some,pichler2022risk}. Threshold policies, such as $k$-search, relate decisions directly to observed prices, while guaranteeing worst-case coverage, and allow for identifying a stopping time criterion. Recent works have extended online $k$‐search methods specifically to the energy‐storage arbitrage problem. The authors in \cite{lorenz2009optimal} derive $k$‐threshold purchase rules that guarantee a competitive ratio against a clairvoyant (perfect foresight) benchmark. The authors in \cite{yang2020online} develop a unified online framework in which both buying and selling thresholds are dynamically adjusted based on the remaining action budget, while \cite{zhang2021arbitrage} analyzes arbitrage opportunities in different storage technologies and shows that simple threshold policies achieve near‐optimal performance under realistic price distributions. These rules are attractive because each threshold has a clear and traceable operational interpretation for the system operator, unlike a learned controller's decisions. Our thresholds are moreover state- and forecast-conditioned rather than static, as they depend on ($t$,$e_t$), and the forecast price bounds. They remain compatible with learned forecasts \cite{lee2024online,sang2022electricity} or calibrated prediction sets \cite{yeh2024end,wu2025online} without changing the threshold mechanics or the competitive-ratio guarantee. Existing $k$-search formulations for storage \cite{lee2024online}, however, rely on thresholds that are static in time and agnostic to state feasibility, and none link these resulting decisions to a terminal state reachability guarantee, which are the two gaps this paper addresses.

\emph{(iv) Uncertainty quantification for storage and inventory.} Recent machine learning approaches to inventory management and energy storage under price uncertainty increasingly emphasize reliable uncertainty quantification rather than point forecasting alone. Some works integrate learned forecasts into online policies \cite{lee2024online} or thresholds to guide real time decisions \cite{sang2022electricity}, while others incorporate switching costs into online optimization via threshold-based designs \cite{lechowicz2023online, li2020online}. A key recent direction is to couple decision-making with calibrated prediction sets as in \cite{yeh2024end, wu2025online} to produce finite-sample prediction intervals for policy outcomes and to provide distribution coverage under data exchangeability assumptions. In our setting, conformal calibration provides coverage-guaranteed bounds for the terminal SoC as a distribution-free uncertainty layer. We empirically compare probability estimates based on assumed price distributions against conformalized quantile regression. To our knowledge, no prior work certifies reachability of a terminal SoC target.

\subsection{Contributions}
The goal of this paper is to propose an operational framework that reconciles asset-level arbitrage decisions with system-level reliability requirements by quantifying the probability that a battery's SoC satisfies an end-of-horizon SoC target band constraint. Our contributions are:
\begin{itemize}
    \item A computationally tractable framework for estimating the probability that a battery's terminal SoC reaches a prescribed target band $\mathcal{E}$. The framework extends the $k$-search method of \cite{lorenz2009optimal} using time-dependent thresholds conditioned on the remaining feasible battery actions. Unlike the static thresholds in \cite{lorenz2009optimal,yang2020online,zhang2021arbitrage}, the proposed thresholds link arbitrage decisions directly to terminal reachability.
    \item An exact pruning and probability-redistribution algorithm that computes the terminal SoC distribution under capacity and rate limits and enables the stopping time $\tau^{\star}$, providing system operators with advance warning of compliance risk. This guarantee is not provided by either the ex-post valuation and market-operation studies, e.g., \cite{wu2024market}, nor by learned storage controllers, e.g., \cite{wang2018energy}.
    \item An empirical validation using six years of real PJM day-ahead price data, comparing the assumed-distribution probability estimates with a distribution-free conformalized quantile regression baseline. The analysis includes a pruning ablation and an out-of-sample evaluation of when each probability estimate remains reliable.
\end{itemize}

\section{Energy arbitrage under uncertain prices} \label{sec:formulation}
Consider the battery arbitrage problem, where the battery operator (BO) aims to maximize its energy arbitrage profits under uncertain energy prices ($\boldsymbol{\lambda}_t$) by charging ($c_t$) when the price is low, and discharging ($d_t$) when it is high. The price is revealed sequentially over the horizon, and the BO makes its operating decisions before observing the realization of $\boldsymbol{\lambda}_t$. Therefore, the BO must account for the estimation of future energy values when determining its current decisions. The battery arbitrage problem can be formulated as follows:
\begin{subequations} \label{P0}
\begin{align} 
    \max_{e,c,d}~&\mathbb{E}_{\boldsymbol{\lambda}} \Big[\sum_{t=1}^T \boldsymbol{\lambda}_t (d_t - c_t) \Big] \label{P0:objective_function}\\
    \text{s.t.}~& \Big\{E_{\min} \leq e_t \leq E_{\max}, &&  \label{P0:storage_capacity}\\
    &e_t ~= e_{t-1} - \tfrac{d_t}{\eta} + c_t \eta, && \label{P0:battery_dynamics} \\
    &0 \leq c_t \leq P,~ 0 \leq d_t \leq P,\Big\}, && \forall \; t \in \mathcal{T}, \label{P0:charge_discharge_limit}
\end{align}    
\end{subequations}
where eq. \eqref{P0:objective_function} is the BO objective function that aims to maximize its profit from energy arbitrage under price uncertainty. Eq. \eqref{P0:storage_capacity} denotes the battery's capacity limits. Constraint \eqref{P0:battery_dynamics} represents the SoC dynamics of the battery, where $\eta\leq1$ is used to denote the efficiency of charging and discharging. Finally, constraints in \eqref{P0:charge_discharge_limit} represent the maximum charging and discharging rate limits ($P$), respectively. 

As mentioned in \cite{tapia2027learning}, problem \eqref{P0} can be formulated as a dynamic optimization problem that incorporates an additional penalty factor to enforce a prescribed SoC target band ($e^{\rm target}$). Rather than imposing a penalty cost relative to the difference between an attained terminal SoC value and $e^{\rm target}$, the system operator (SO) could enforce a minimum requirement through a chance constraint. Specifically, at the terminal time $T$, the SO requires the SoC to stay within a tolerance band $\delta$ around $e^{\rm target}$ for at least $(1\!-\!\epsilon)$ of price realizations. This constraint defines a security margin with a user-defined risk tolerance. Thus, for the prescribed SoC target band $\mathcal{E}=[e^{\text{target}} \pm \delta]$ for the time $ \tau \in \mathcal{T}^{\text{target}}$, the chance constraint is defined as follows: 
\begin{equation}
    \mathbb{P}_{\boldsymbol{\lambda}}\{ e_{\hat{\tau}} \in \mathcal{E} \} \; \geq (1\!-\!\epsilon), \quad \forall \hat{\tau} \in \mathcal{T}^{\text{target}}. \label{prob_CC}
\end{equation}
\noindent Since $d_t$ and $c_t$ are continuous variables, the future SoC admits infinitely many realizations, making its distribution intractable. To address this issue, we consider a fixed rate ($P$) for charging and discharging. Let $a_t$ represent the BO action at time $t$ that replaces the variables $c_t$ and $d_t$. Then, $a_t$ takes the value: $+1$ for charging, $-1$ for discharging, and $0$ for idle. To facilitate the analysis, we consider a single target time, $\mathcal{T}^{\text{target}} = \{T\}$, and perfect battery efficiency ($\eta = 1$), although more general efficiency models can be incorporated, for example, by appropriately adjusting the charging and discharging rates. Under these assumptions, the battery arbitrage problem is equivalently stated as follows: 
\begin{subequations}
\begin{align}
    \max_{a, e} \quad & \mathbb{E}_{\boldsymbol{\lambda}} \Big[ \sum_{t =1}^{T} - \boldsymbol{\lambda}_{t}\,a_{t} \,P \label{P1:objective_function}\Big] \\
    \text{s.t.} \quad & E_{\min} \leq e_t \leq E_{\max}, && \forall t \in \mathcal{T} \nonumber\\
    & e_{t} = e_{t-1} + a_{t}\,P , && \forall t \in \mathcal{T} \label{P1:battery_dynamics}\\
    & a_{t} = \left\{\begin{array}{lr}
        1, & \text{if charge }\\
        -1, & \text{if discharge}\\
        0, & \text{if idle} \\
        \end{array}\right\}, && \forall t \in \mathcal{T} \label{P1:battery_actions}\\
    & \mathbb{P}_{\boldsymbol{\lambda}}\{e_{T} \in \mathcal{E}\} \geq (1\!-\!\epsilon), \label{P1:chance_constraint}
\end{align}    
\end{subequations}
where the expression in \eqref{P1:objective_function} corresponds to the maximization of the battery arbitrage profit considering a fixed charging and discharging rate ($P$). Eq. \eqref{P1:battery_dynamics} corresponds to the battery SoC dynamics and ensures that changes in the SoC match the BO actions ($a_t$) multiplied by $P$. The constraint in \eqref{P1:battery_actions} restricts control of the actions by the BO, and Eq. \eqref{P1:chance_constraint} is a chance constraint designed to ensure that the terminal SoC ($e_T$) meets the prescribed target band with a probability at least $(1\!-\!\epsilon)$, where $\epsilon$ is a user-defined risk tolerance.

As a result, problem \eqref{P0} reduces to a tractable special case that admits the closed-loop policy for the price threshold analysis, that we derived in Sec. \ref{subsection_threshold}, at the expense of no longer capturing continuous dispatch.

\subsection{Price thresholds and probability transitions estimate} \label{subsection_threshold}
Threshold policies are decision rules based on predefined cut-off values and available information to support real time decision-making. These methods are particularly useful in inventory settings because of their simple interpretability (since they can be implemented using only the current system state, and each threshold has a direct operational meaning). They are also aligned with an energy arbitrage problem, which requires making sequential decisions under uncertainty as market prices are revealed over time. These algorithms design a threshold function that defines a minimum (or maximum) acceptable value, or a sequence of such values, that each newly revealed price must satisfy for the BO to take an action. These threshold policies determine the price conditions under which an action is taken and, consequently, assign an economic value to the stopping criterion, i.e. when no further action is taken.

\begin{figure}[ht]
    \centering
    \includegraphics[width=0.9\linewidth, trim={0 2mm 0 2mm},clip]{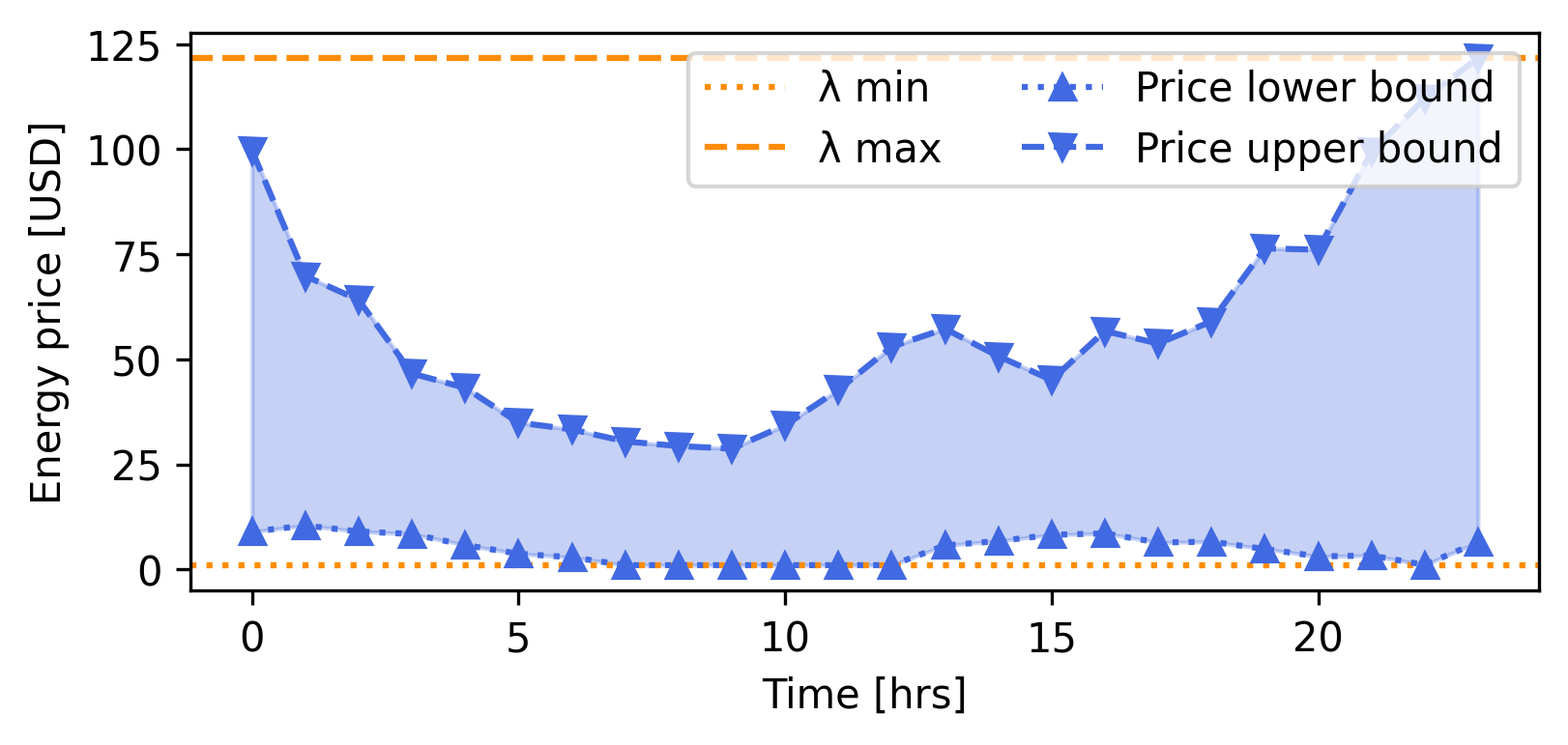}
    \caption{Full-day price uncertainty set based on the PJM real time prices data from 2011-2016. Dashed orange lines represent the minimum ($\lambda^{\min}$) and maximum ($\lambda^{\max}$) price values across the time horizon, while the dashed blue lines bound a possible realization for the energy price ($\boldsymbol{\lambda}_t$).}
    \label{fig:multi_band}
\end{figure}

In the context of battery arbitrage, the BO commits an energy schedule (i.e. a SoC trajectory) in the DA market. In RT market, BO actions may differ from this schedule, and the resulting deviations are settled at the realized RT price. Accordingly, the BO must decide whether to buy (charge), sell (discharge) or idle (stick to the DA SoC trajectory) using only the current observed RT price, without knowing future price realizations or deviations from the DA schedule. To account for uncertainty realizations, BOs often rely on scenario generation, robust price sets, or probability distributions \cite{sioshansi2009estimating}. Fig. \ref{fig:multi_band} shows a real-world example of a full-day electricity price uncertainty set.

To evaluate the performance of a given policy in the absence of probabilistic models, the \emph{competitive ratio} is commonly used, which corresponds to the worst‐case ratio between the outcome of an online algorithm policy $\rm ALG_{\pi}(\lambda)$ and the offline optimum $\rm OPT(\lambda)$ with full knowledge of the future. For instance, $k$-search algorithms are online methods that use threshold rules to make sequential decisions under uncertainty and have competitive guarantees. For a maximization problem, the competitive ratio $\alpha\! =\! \sup_{\lambda} \frac{\rm OPT(\lambda)}{\rm ALG_{\pi}(\lambda)} \geq 1$. Equivalently, the policy $\pi$ is $\alpha$-competitive if $\mathrm{ALG}_{\pi}(\lambda)\geq \frac{1}{\alpha}\mathrm{OPT}(\lambda)$ for all $\lambda$, i.e., the online policy is guaranteed to achieve at least a $1/\alpha$ fraction of the offline optimum in the worst case. For a minimization problem, the definition of the competitive ratio $\omega$ is typically inverted \cite{lorenz2009optimal}. We use this convention for $\alpha \geq 1$ and $\omega \geq 1$ throughout the rest of the theoretical development of the threshold policy.

When the values $\lambda^{\min}$ and $\lambda^{\max}$ are available, $k$‐search algorithms leverage these bounds to construct a sequence of thresholds for \emph{buy} and \emph{sell} decisions. In a $k$‐min search, the algorithm allows up to $k_{\rm ch}$ purchases at precomputed \emph{buy} thresholds, whereas a $k$‐max search allows up to $k_{\rm dis}$ sales when prices exceed \emph{sell} thresholds. These thresholds are geometrically spaced across the price range to guarantee a provable competitive ratio relative to the offline optimal result under perfect foresight \cite{lorenz2009optimal}. 

We assume the BO adopts a pre-policy ($k_{\rm dis}$,$k_{\rm ch}$), i.e., the policy obtained before pruning in Sec. \ref{subsection:pruning}, that indicates the predefined number of possible discharge ($k_{\rm dis}$) and charge ($k_{\rm ch}$) actions. Then, we can use the $k$-search algorithm to derive price thresholds to support the decision process of the BO. Thus, based on Alg. 1 in \cite{sun2020competitive} the $j-$th threshold to buy (charge) is defined as follows:
\begin{align} \label{threshold_u}
u_j^{\rm ch} & = \lambda^{\max} \left[ 1 - \left( 1 - \tfrac{1}{\alpha} \right) \left( 1 + \tfrac{1}{\alpha k_{\rm ch}} \right)^{j-1} \right],
\end{align}
and $i-$th threshold to sell (discharge) energy corresponds to:
\begin{align} \label{threshold_l}
\ell_i^{\rm dis} = \lambda^{\min} \left[ 1 + (\omega - 1) \left( 1 + \tfrac{\omega}{k_{\rm dis}} \right)^{i-1} \right],
\end{align}
where $j\!\in\![1,k_{\rm ch}]$ and $i\!\in\![1,k_{\rm dis}]$, and $\alpha$ and $\omega$ are the competitive ratios of the algorithm for charging and discharging, respectively. The thresholds in \eqref{threshold_u} and \eqref{threshold_l} do not consider the battery capacity limits but, as we show later, can be handled using explicit pruning and renormalization techniques described in Section \ref{subsection:pruning}.


In the worst-case setting, $k_{\rm ch}$ and $k_{\rm dis}$ can be tuned to achieve a desired trade-off between aggressiveness in purchasing/selling and conservatism in preserving the remaining budget. Given these parameters, the reservation price policy (i.e., the maximum price of charging or the minimum price of discharging) is defined by these two sequences of the charging thresholds $\{u^{\rm ch}_j\}_{j=1}^{k_{\rm ch}}$ and the discharging thresholds $\{\ell^{\rm dis}_i\}_{i=1}^{k_{\rm dis}}$. At each time $t$, the policy observes the revealed price $\lambda_t$ and compares it with the next available threshold. If fewer than $k_{\rm ch}$ charges have been executed, $\lambda_t\!\le\!u^{\rm ch}_j$, and $e_{t-1}\!+\!P\!\leq\!E_{\max}$, the BO charges. Otherwise, if fewer than $k_{\rm dis}$ discharges have been executed, $\lambda_t\!\ge\!\ell^{\rm dis}_{i}$, and $e_{t-1}\!-\!P\!\geq\!E_{\min}$, the BO discharges. If neither condition is met, the battery remains idle. If the realized price crosses a threshold but the corresponding action violates the capacity limits, the action is treated as idle and the associated threshold counter ($i$ or $j$) is not advanced.

In practice, asset operators manage market volatility using time-varying probabilistic price bounds or calibrated predictive models rather than deterministic point predictions \cite{nowotarski2018recent}. These bounds are often supplied by commercial forecasting platforms (e.g., \cite{gridmatic_2026}, \cite{gridstatus_2026}). Accordingly, we assume that the BO has access to time-varying price bounds over the remaining horizon as illustrated by the blue lines in Fig. \ref{fig:multi_band}. Note that tighter price bounds improve the competitive ratio but increase the risk of prices falling outside the range, reducing profit. Thus, bound width trades off guarantees and economic performance. Sec. \ref{section_CQR} introduces a distribution-free CQR alternative. At time $t$, these bounds are denoted by $z^{\rm min}(t)\!=\!\{\lambda^{\min}_{s}\}_{s=t}^{T}$ and $z^{\rm max}(t)\!=\!\{\lambda^{\max}_{s}\}_{s=t}^{T}$. As time advances and the remaining horizon shrinks, these bounds are updated accordingly. We denote by $z^{\rm min}(t)$ the non-decreasing ordered list of the remaining $m$ lower bounds at time $t$ such that $z_1^{\rm min}(t)\!\le\!\cdots\le\! z_m^{\rm min}(t)\!$, and by $z^{\rm max}(t)$ the non-increasing ordered list of the remaining $m$ upper bounds at time $t$ such that $z_1^{\rm max}(t)\!\ge\!\cdots\!\ge\! z_m^{\rm max}(t)$. Using the information revealed up to $t$, we adjust the $j$-th price threshold for the $k$-min algorithm to maintain the $\alpha$-competitive guarantee as:
\begin{equation}
    u_j^{\rm ch}(t) = \frac{1}{\alpha k_{\rm ch}} \Big( \sum_{\hat{t}=1}^{t-1} \sum_{k=1}^{j-1} u_k^{*\rm ch}(\hat{t}) + \sum_{n=1}^{k_{\rm ch}-(j-1)} z^{\rm max}_n(t) \Big).\label{eq_ex_1}
\end{equation}
Similarly, we modify the $i$-th price threshold for the $k$-max algorithm, maintaining the $\omega$ ratio:
\begin{equation}
    \ell_i^{\rm dis}(t) = \frac{\omega}{k_{\rm dis}} \Big( \sum_{\hat{t}=1}^{t-1} \sum_{k=1}^{i-1} \ell_k^{*\rm dis}(\hat{t}) + \sum_{n=1}^{k_{\rm dis}-(i-1)} z^{\rm min}_n(t) \Big). \label{eq_ex_2}
\end{equation}
Here, the first term accounts for previously activated price thresholds and the related corresponding actions. In particular, $u_k^{*\rm ch}(\hat{t})$ in \eqref{eq_ex_1} and $\ell_k^{*\rm dis}(\hat{t})$ in \eqref{eq_ex_2} represent the charging and discharging thresholds activated at some time $\hat{t}<t$. The second term in \eqref{eq_ex_1} and \eqref{eq_ex_2} sums the remaining worst-case values in the ordered lists $z^{\rm max}(t)$ and $z^{\rm min}(t)$, respectively. Together, these terms adjust the price threshold to avoid underestimating the set of feasible worst-case realizations, since some trajectories within the bounds $[\lambda^{\rm min}, \lambda^{\rm max}]$ may no longer be reachable given the remaining time and SoC constraints. To exclude infeasible decision trajectories at $(t,e_t)$, we impose the following time- and state-dependent bounds:
\begin{equation*}
    \{z^{\rm min}_n (t)\}_{n=1}^{k_{\rm dis} - (i-1)} \rightarrow \{z^{\rm min|feas}_n (t, e_t)\}_{n=1}^{K^{\rm feas}_{\rm min}(t,e_t)}.
\end{equation*}
Thus, by considering only feasible SoC trajectories, the $i$-th price threshold for the $k$-max algorithm is defined as follows:
\begin{align}
\ell_{i}^{\rm dis}(t,\!e_t)\!=\!\frac{\omega}{K^{\rm feas}_{\rm min}(t,\!e_t)} \Big(\!\sum_{\hat{t}=1}^{t-1} \sum_{k=1}^{i-1} \ell_k^{\rm dis \star}(\hat{t})\!
+\!\!\!\!\!\!\!\!\sum_{n=1}^{K^{\rm feas}_{\rm min}(t,e_t)} \!\!\!\!\!z_{n}^{\rm min|feas}(t,\!e_t)\!\Big), \label{NewThreshold}
\end{align}
where $K_{\rm min}^{\rm feas}(t, e_t)$ denotes the maximum number of discharge actions feasible from the current SoC at $t$, and $z_{n}^{\rm min|feas}(t,e_t)$ is the $n$-th smallest predicted price restricted to future time slots reachable by at least one feasible discharge trajectory. The same logic is used to compute the charging thresholds ($u_{i}^{\rm ch}(t, e_t)$) by restricting the corresponding price sets to feasible trajectories. Appendix \ref{app:feasible_pruned} details how $K^{\rm feas}(t,e_t)$ is computed from the SoC and rate limits, as well as the construction of the feasibility-conditioned predicted price sets $z_{n}^{\rm min|feas}(t,e_t)$ used in the benchmark. Because the time-dependent thresholds are recomputed at every $(t,e_t)$ using only the feasible remainder of the price bound sequence, the decision at time $t$ already reflects the battery's full operating history through $e_t$ and the exhausted threshold budget $(k_{\rm ch},k_{\rm dis})$. This is what allows the framework to preserve a competitive-ratio guarantee (Appendix A) and a chance-constraint quantification simultaneously, a combination not available, to our knowledge, for learned policies.

Thus, by having access to time-varying price bounds, the BO can leverage forecast information in its decision-making process. Given the predetermined threshold sequences $\{u_j(t)\}_{j=1}^{k_{\rm ch}}$ and $\{\ell_i(t)\}_{i=1}^{k_{\rm dis}}$, the BO can use the price distribution $F(\boldsymbol{\lambda})$ to compute at each time $t$ the probabilities of charging, discharging, or idling. We denote these probabilities by $\alpha_t$, $\beta_t$, and $\gamma_t$, respectively, and use them to evaluate the chance constraint in \eqref{P1:chance_constraint}. For instance,
\begin{subequations}\label{probs}
    \begin{align} 
        \alpha_t \!=\!\!\!\int \!\! F \Big( \boldsymbol{\lambda}_t \leq u^{\rm ch}(t)\! \Big) d\boldsymbol{\lambda}_t,\\
        \beta_t\!=\!\!\!\int \!\! F \Big(\boldsymbol{\lambda}_t \geq \ell^{\rm dis}(t)\!\Big) d\boldsymbol{\lambda}_t,\\
        \gamma_t\! =\! 1\! -\! \alpha_t\! -\! \beta_t.
    \end{align}
\end{subequations}

\noindent Expressions in \eqref{probs} are aligned with the decision rules described in \eqref{eq_ex_1} and \eqref{eq_ex_2}, and give the time dependent probabilities of charging, discharging, and idling under $F(\boldsymbol{\lambda})$. Propagating these probabilities through the SoC constraint in \eqref{P1:battery_dynamics} yields the induced distribution of $e_T$, which allows us to evaluate \eqref{P1:chance_constraint}. 

\subsection{Uncertainty quantification for the terminal state-of-charge} \label{subsection:uncertainty}
Under the threshold price policy described in Sec. \ref{subsection_threshold}, we assume that the probability of taking action $a_t$ can be measured by evaluating a possible price distribution of $\boldsymbol{\lambda}_t$. The probabilities $\alpha_t, \beta_t, \gamma_t \geq 0$ from \eqref{probs} satisfy $\alpha_t + \beta_t + \gamma_t = 1$, forming a valid distribution over the three mutually exclusive actions. Hence, the terminal SoC can be quantified by considering $e_T\!=\!e_0\!+\!\sum_{t=1}^{T} Y_t$, where $Y_t\!=\!a_t\,P$ is the net injection at $t$, which is defined as:
\begin{equation}
Y_t\!=\!a_tP\!=\!
\begin{cases}
+P,\,0,\,-P
\end{cases}
\text{w.p. } \alpha_t,\;\gamma_t,\;\beta_t, \text{ resp.}
\end{equation}
We are interested in computing $\mathbb{P}_{\boldsymbol{\lambda}}\{ e_{T} \!\in\! \mathcal{E} \}$, and then, in estimating the number of charging ($r_{\rm ch}$) and discharging ($r_{\rm dis}$) actions to compute $e_T$. Consider then, 
$r_{\mathrm{ch}}\!=\!\sum_{t=1}^{T}\mathbbm{1}(a_t\!=\!1), \, r_{\mathrm{dis}}\!=\!\sum_{t=1}^{T}\mathbbm{1}(a_t\!=\!-1),$
where $\mathbbm{1}(\cdot)$ corresponds to an indicator function, equal to $1$ when the condition inside the parentheses holds, and $0$ otherwise. The terminal SoC satisfies $e_T\!=\!e_0\!+\!P\,\bigl(r_{\rm ch}\!-\! r_{\rm dis}\bigr)$. Therefore, the set of trajectories that lead to $e_T \in \mathcal{E}$ can be characterized in terms of the bounds on the net number of charging and discharging actions, i.e. $\underline{S}\!\le\! r_{\mathrm{ch}}\!-\!r_{\mathrm{dis}}\!\le\! \overline{S}$ where $\underline{S}$ and $\overline{S}$ are defined as follows:
\begin{equation}
    \underline{S} = \tfrac{ \lceil e^{\rm target}-\delta - e_0\rceil}{P}, \qquad \overline{S} = \tfrac{\lfloor e^{\rm target}+\delta - e_0\rfloor}{P}. \label{prev_eq} 
\end{equation}
The probability of finishing within the SoC target band is:
\begin{equation*}
    \mathbb{P}_{\boldsymbol{\lambda}}\{e_{T} \in \mathcal{E} \} = \mathbb{P}_{\boldsymbol{\lambda}}\{ \underline{S} \;\le\; r_{\mathrm{ch}} - r_{\mathrm{dis}} \;\le\; \overline{S}\},
\end{equation*}
which then can be expressed using \eqref{probs} and \eqref{prev_eq} as follows:
\begin{subequations} \label{prob_target_band}
    \begin{align}
    \mathbb{P}_{\boldsymbol{\lambda}}\{e_{T} \in \mathcal{E}\;|\;\kappa_{k_{\rm dis}}^{k_{\rm ch}}\} = \qquad \qquad \qquad \qquad \qquad \qquad \nonumber \\ 
    \kappa_{k_{\rm dis}}^{k_{\rm ch}} \sum_{\Delta r}
    \;\sum_{\substack{A\subseteq\mathcal{T}}}
    \: \sum_{\substack{B\subseteq\mathcal{T}\setminus A}}
    \;\prod_{t\in A}\alpha_t\, \prod_{t\in B}\beta_t\, \prod_{t\notin A\cup B}\gamma_t, \label{product} \\
    \substack{\underline{S} \le \Delta r = r_{\mathrm{ch}}-r_{\mathrm{dis}} \le \overline{S}},\quad|A|=r_{\mathrm{ch}}, \quad |B|=r_{\mathrm{dis}},
    \end{align}   
\end{subequations}
where $\kappa_{k_{\rm dis}}^{k_{\rm ch}}$ corresponds to the probability that the BO chooses a specific pre-policy ($k_{\rm ch}$,$k_{\rm dis}$), and $\sum_{k_{\rm dis}} \sum_{k_{\rm ch}} \kappa_{k_{\rm dis}}^{k_{\rm ch}}\!=\! 1$ . Here, $A\!=\!A(e_t,t)\!\subseteq\!\mathcal{T}$ is the subset of time indices in which the BO charges, $A(e_t,t)\!=\!\{t\!\in\!\mathcal{T} \!\colon\! \mathbbm{1}(a_t\!=\!1)\!=\!1, \, e_t\!\in\![E_{\rm min},E_{\rm max}] \,\forall t\!\le\!T\}$ and $B\!=\!B(e_t,t)\!\subseteq\!\mathcal{T}\!\setminus\!A$ is the subset of time indices in which the BO discharges, $B(e_t,t)\!=\!\{t\!\in\!\mathcal{T}\!\colon\! \mathbbm{1}(a_t\!=\!-1)\!=\!1, \,e_t\!\in\![E_{\rm min},E_{\rm max}] \,\forall t\le\!T\}$. The product term in \eqref{product} depends only on $t$, which is exact conditional on the trajectory staying feasible and the $(k_{\rm ch},k_{\rm dis})$ budget not being exhausted. It does not account for trajectories where $e_t\!\notin\![E_{\rm min},E_{\rm max}]$ or that exhaust the ($k_{\rm ch}, k_{\rm dis}$) budget. To evaluate \eqref{P1:chance_constraint}, we employ the pruning and probability redistribution algorithm in Sec. \ref{subsection:pruning}, which enumerates admissible action sequences, removes infeasible trajectories, and redistributes the associated probability mass among the remaining feasible sequences. By evaluating $\alpha_t, \beta_t, \gamma_t$ along each previously realized sequence, this procedure yields the exact $e_T$ distribution conditioned on feasibility, formalized as \emph{feasibility-conditioned renormalization} in Sec. \ref{subsection:pruning}.

We assume the BO enumerates candidate pre-policy tuples $(k_{\rm ch},k_{\rm dis})$ and computes the expected profit of each based on the price thresholds and the assumed price distribution.
If the tuple is not profitable, it is discarded. Among the remaining combinations, the probability of selecting any one of them is uniform across them. Finally, if the remaining horizon or the current SoC can no longer support the originally planned $(k_{\rm ch}, k_{\rm dis})$ actions given the realized price sequence $\boldsymbol{\lambda}$, the corresponding values are reduced to preserve feasibility over the remaining time, as described in Appendix \ref{app:feasible_pruned}.

\subsection{Pruning and probability redistribution algorithm} \label{subsection:pruning}
\begin{figure}[ht]
    \centering
    \includegraphics[width=1.05\linewidth, trim={0mm 0mm 0mm 0mm},clip]{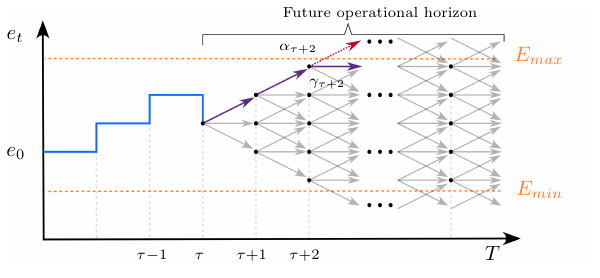}
    \caption{Feasibility-conditioned renormalization. The purple path shows one trajectory from $\tau$ to $\tau\!+\!2$. At $\tau\!+\!2$, charging is infeasible (red arrow), so its probability $\alpha_{\tau+2}$ is reassigned to the idle action, yielding $\gamma_{\tau+2}\leftarrow \gamma_{\tau+2}+\alpha_{\tau+2}$ and $\alpha_{\tau+2}\leftarrow 0$.}
    \label{fig:enter-label3}
\end{figure}

To compute the exact SoC distribution under operational constraints, we propose a pruning approach, which we call \emph{feasibility-conditioned renormalization}. For a fixed pre-policy $(k_{\rm ch}, k_{\rm dis})$, we define $\mathcal{D}\! =\! \{-1,0,1\}^{|\mathcal{T}|}$ as the set of all action sequences $d\! =\! \{a_t\}_{t=1}^{T}$. Each sequence has a raw unconstrained probability:
\begin{equation}
    P^{\dagger}(d)\!=\! \prod_{t \in \mathcal{T}} \pi_{t}(a_t), \; \text{with } \pi_{t}(a_t) = 
    \begin{cases}
    \alpha_t, & \text{if } a_t = 1,\\
    \gamma_t, & \text{if } a_t = 0,\\
    \beta_t, & \text{if } a_t = -1.
    \end{cases}
    \label{prob_raw}
\end{equation}
where probabilities are evaluated using the thresholds in Sec. \ref{subsection_threshold} along each previously realized sequence, so no two distinct sequences are collapsed into a shared state. Let $\mathcal{D}^{\rm feas} \!\subseteq\! \mathcal{D}$ be the subset of trajectories that satisfy $e_t\! \in\! [E_{\rm min},E_{\rm max}]\,\forall t$. We prune infeasible sequences in $\mathcal{D}\! \setminus\! \mathcal{D}^{\rm feas}$ and renormalize the remaining probability mass as:
\begin{equation}
    \mathbb{P}_{\boldsymbol{\lambda}}\{d|\mathcal{D}^{\rm feas}\} = \frac{P^{\dagger}(d) \, \mathbbm{1}(d\in\mathcal{D}^{\rm feas})}{\sum_{d' \in \mathcal{D}^{\rm feas}} P^{\dagger}(d')}. \label{eq_renorm}
\end{equation}
We define $\mathrm{dist}_T[e] = \sum_{d\in\mathcal{D}^{\rm feas}:\, e_T(d)=e} \mathbb{P}_{\boldsymbol{\lambda}}\{d\mid\mathcal{D}^{\rm feas}\}$, from which we compute the probability of meeting the terminal band and the stopping time. Accordingly, \eqref{eq_renorm} conditions on the event that the \emph{entire} sequence remains within $[E_{\rm min},E_{\rm max}]$. Under forward propagation, the probability of each infeasible action at a given state is reassigned to the idle branch ($\gamma_t\!\leftarrow\!\gamma_t+\alpha_t$ or $\gamma_t\!\leftarrow\!\gamma_t+\beta_t$, with the infeasible mass then set to zero) before propagating from $t\!=\!1$ to $T$. Because \eqref{eq_renorm} only reweights already-decided sequences and never alters the online decision rule $\mathrm{ALG}_\pi$ or the thresholds that generate them, the competitive-ratio guarantee of Appendix \ref{app:proofs} is unaffected by pruning. In Appendix \ref{app:mid_horizon_k} we discuss how the budget for ($k_{\rm ch},k_{\rm dis}$) is handled when a sequence exhausts its threshold budget before $T$. Fig. \ref{fig:enter-label3} shows an example of how the feasibility-conditioned renormalization algorithm works.

\begin{figure}[ht]
  \centering
  \begin{subfigure}[t]{0.24\textwidth}
    \centering
    \includegraphics[width=1.03\textwidth, trim={0 0mm 0 0mm},clip]{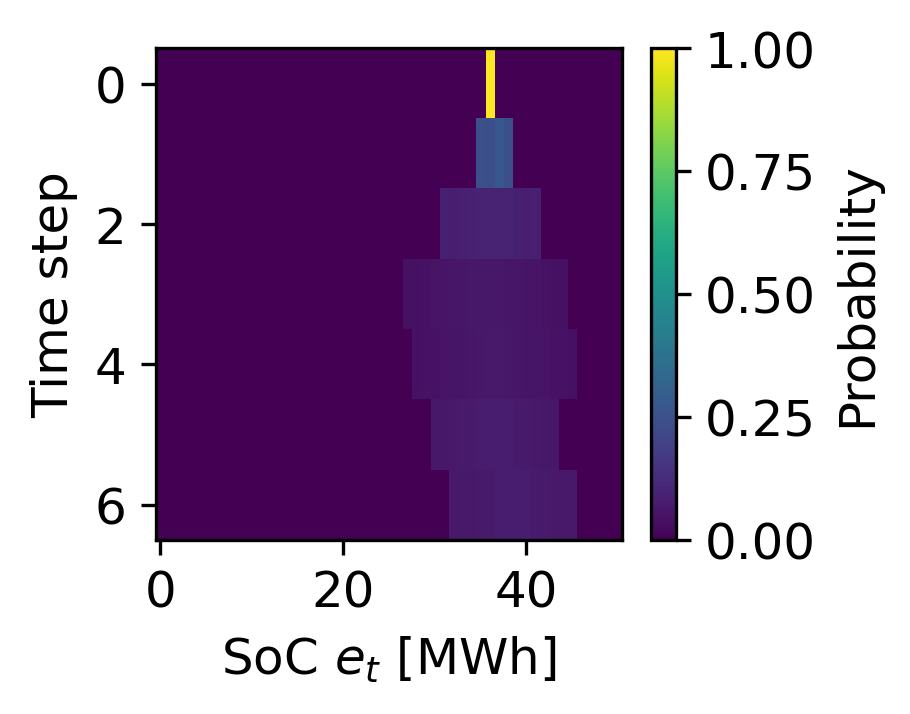}
    \caption{Non-pruned SoC probability distribution over time}
    \label{fig:foobar_a}
  \end{subfigure}
  \hfill
  \begin{subfigure}[t]{0.24\textwidth}
    \centering
    \includegraphics[width=1.03\textwidth, trim={0 0mm 0 0mm}, clip]{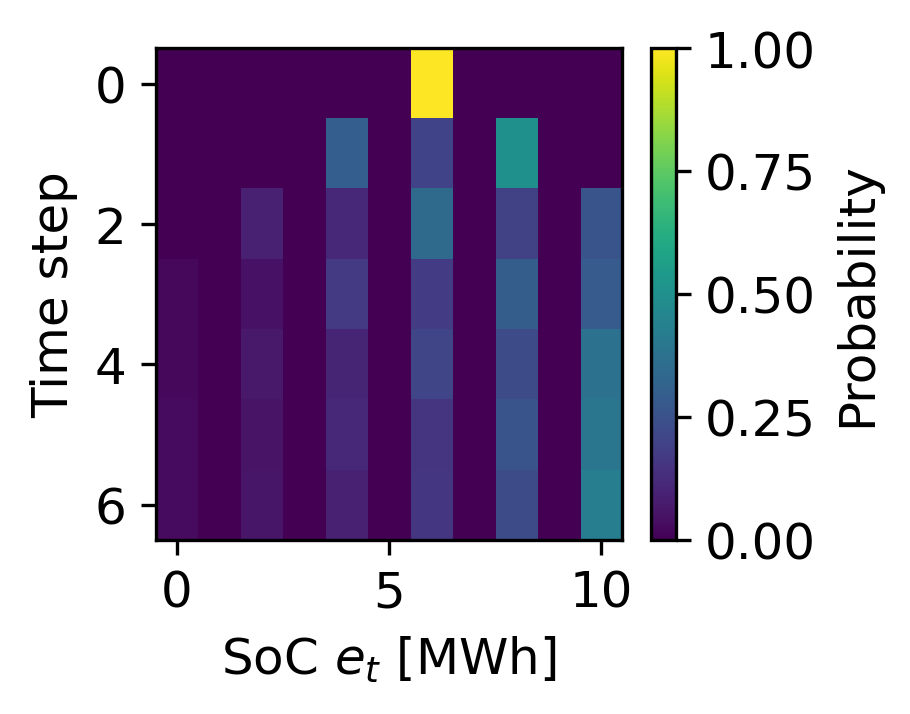}
    \caption{Pruned SoC probability distribution over time}
    \label{fig:foobar_b}
  \end{subfigure}
  \hfill
  \begin{subfigure}[t]{0.24\textwidth}
      \centering
      \includegraphics[width=\textwidth, trim={0 0mm 0 0mm}, clip]{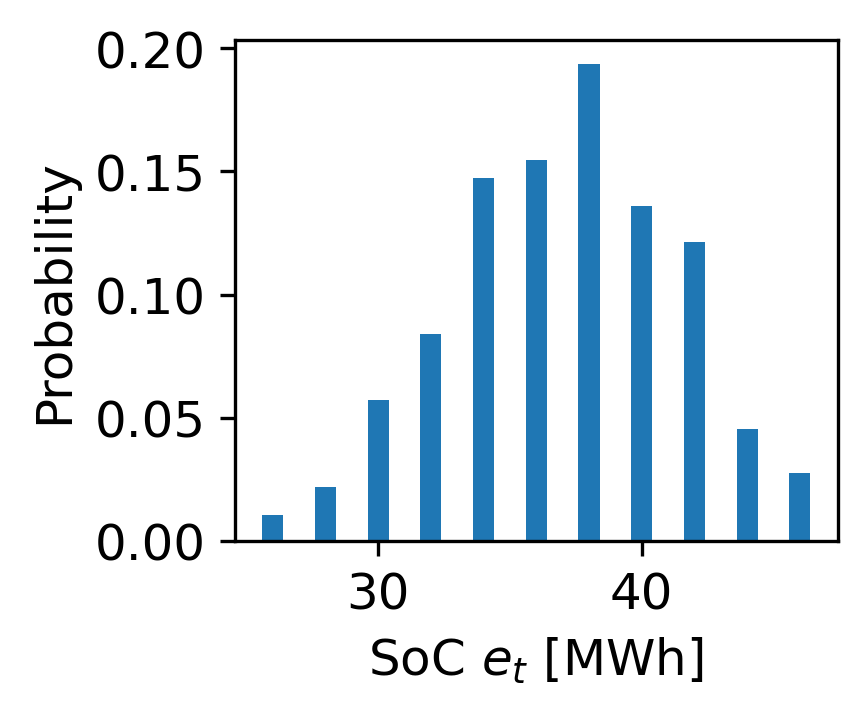}
      \caption{Non-pruned terminal SoC probability distribution}
      \label{fig:foobar_c}
   \end{subfigure}
  \hfill
  \begin{subfigure}[t]{0.24\textwidth}
      \centering
      \includegraphics[width=\textwidth, trim={0 0mm 0 0mm}, clip]{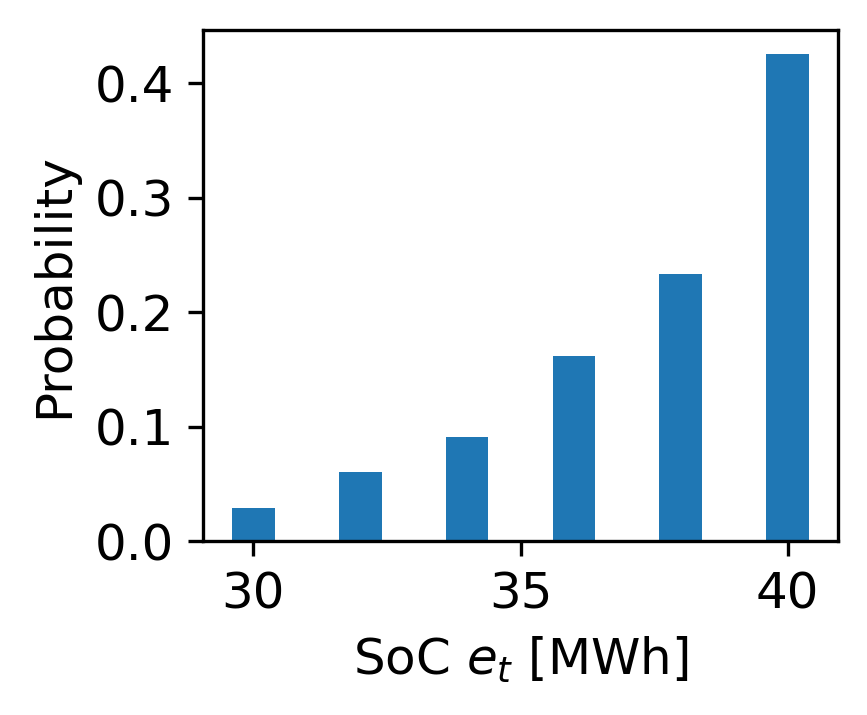}
      \caption{Pruned terminal SoC probability distribution}
      \label{fig:foobar_d}
   \end{subfigure}
  \caption{(a) Non-pruning SoC distribution over time. (b) Pruning SoC distribution over time. (c) Non-pruning terminal SoC distribution. (d) Pruning terminal SoC distribution.}
  \label{fig:foobar}
\end{figure}

\noindent Fig. \ref{fig:foobar} shows a probability heat map for an illustrative example of SoC trajectories using our proposed algorithm for $T\!=\!6$. Battery limits are set as $E_{\rm min}\!=\!30$ and $E_{\rm max}\!=\!50$ MWh. Fig. \ref{fig:foobar_a} shows that without pruning, the probability mass takes values outside $[E_{\rm min},E_{\rm max}]$ as $t$ increases since \eqref{probs} is evaluated as if actions were always feasible. In contrast, Fig. \ref{fig:foobar_b} prunes sequences that leave $[E_{\rm min}, E_{\rm max}]$ and renormalizes the remaining mass over $\mathcal{D}^{\rm feas}$ via \eqref{eq_renorm}. Figs. \ref{fig:foobar_c} and \ref{fig:foobar_d} show the terminal effect where, without pruning, the probability mass is overstated near the capacity limits, whereas pruning concentrates the feasible probability mass on the boundary states.

\vspace{-1mm}

\subsection{Stopping time for the battery asset}
\begin{figure}[ht]
    \centering
    \vspace{-2mm}
    \includegraphics[width=\linewidth, trim={0mm 0mm 5mm 0mm},clip]{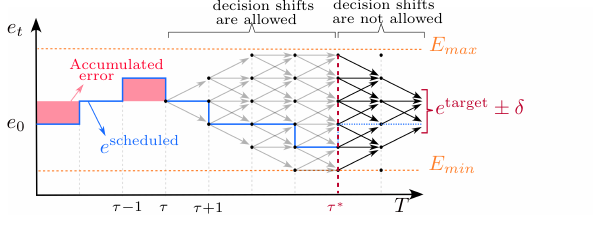}
    \caption{Stopping time interpretation for battery arbitrage. $\tau^{\star}$ is the earliest time at which the chance constraint is already guaranteed under the threshold policy, while balancing arbitrage opportunities against deviations from the DA schedule.}
    \label{fig:enter-label5}
\end{figure}

In online decision problems, stopping time criteria arise from optimal‐stopping theory and the analysis of competitive online algorithms (e.g., the secretary problem). A stopping rule $\tau$ is a stopping time with respect to the observed price process if the decision to stop taking actions at any time $t$ depends only on $\{\lambda_n\}_{n=1}^{t}$. The goal is to design $\tau^{\star}$ to maximize a performance metric, which in this problem corresponds to satisfying the chance constraint in \eqref{P1:chance_constraint}, without foreknowledge of future prices. Thus, the minimum stopping time seeks the earliest period $\tau^{\star}\!\le\!T$ at which $\mathbb{P}_{\boldsymbol{\lambda}}\{e_{T} \in \mathcal{E}\}\! \geq\! (1\!-\!\epsilon)$ is guaranteed under the threshold policy, given the price information $\{\lambda_n\}_{n = 1}^{\tau^{\star}}$ revealed up to $\tau^{\star}$. The existence of such a $\tau^{\star}$ is not guaranteed. If the target reliability becomes unattainable, the SO should be notified sufficiently in advance to allow corrective action. In this case, $\tau^{\star}$ can be interpreted as the \textit{point of no return}, beyond which no further intervention achieves $\mathcal{E}$. Fig. \ref{fig:enter-label5} shows an example of stopping time for the arbitrage problem.

To compute the minimum stopping time, we first evaluate the transition probabilities in \eqref{probs}, and then apply the \emph{feasibility-conditioned renormalization} in Sec. \ref{subsection:pruning} to the full-horizon sequence set $\mathcal{D}^{\rm feas}$ to obtain $\mathrm{dist}_t[e]$ $\forall \, t\in\mathcal{T}$ and feasible $e_t$. For each $(t,e)$, we separately apply the same renormalization to the sub-horizon $\{t+1,\dots,T\}$, restricted to sequences that start from $e_t=e$ with the remaining threshold budget implied by the prior action history, to obtain $\mathbb{P}_{\boldsymbol{\lambda}}\{e_T\!\in\!\mathcal{E},\;e_t\!=\!e\}$. From this, we derive:
\begin{subequations}
\begin{align}
    \mathrm{dist}_t[e]\;=\;\mathbb{P}_{\boldsymbol{\lambda}}\{e_t = e\},\\
    q_t(e) \;=\;\mathbb{P}_{\boldsymbol{\lambda}}\{\,e_T \in \mathcal{E} \mid e_t = e\}.
\end{align}  
\end{subequations}
Then, we compute the marginal probability $\sum_{e}\mathrm{dist}_t[e]\,q_t(e)$ $=\!\mathbb{P}_{\boldsymbol{\lambda}}\{e_T\in\mathcal{E}\}$ $\forall t$, which is constant over $t$ and thus noninformative as a stopping criterion. Instead, $q_t(e_t)$, evaluated at the realized SoC, gives the conditional probability of reaching $\mathcal{E}$ from the current $e_t$. Accordingly, we define the stopping time as $\tau^{\star} \!=\!\min\bigl\{t :q_t(e_t)\! \geq\! (1\! -\!\epsilon)\bigr\}$. Thus, by declaring $\tau^{\star}$ to the SO, the SO gains confidence of at least $(1\!-\!\epsilon)$ that continuing the threshold policy will leave $e_T\!\in\!\mathcal{E}$. Declaring $\tau^{\star}$ ahead of time gives the SO advance notice of when the BO will cease its market participation, which can be incorporated into reserve scheduling, critical-hour planning and other routines. Furthermore, this stopping time lets us derive conservative threshold prices for battery actions that inform the pricing of the stopping action.

Alg. \ref{alg:sop} summarizes the procedure explained in Sections \ref{subsection_threshold} and \ref{subsection:uncertainty}, including the computation of feasibility-aware thresholds, pruning and probability mass redistribution, and stopping time evaluation.
\begin{algorithm}[ht]
\caption{Threshold policy with pruning and stopping time}
\label{alg:sop}
\begin{algorithmic}[1]
\State \textbf{Input:} horizon $\mathcal{T}$, initial SoC $e_0$, limits $E_{\rm min}$ and $E_{\rm max}$, rate $P$, target band $\mathcal{E}$, risk tolerance $\epsilon$, price-bound forecast $\{\lambda^{\min}_\tau,\lambda^{\max}_\tau\}_{\tau\in\mathcal{T}}$, pre-policy ($k_{\rm ch},k_{\rm dis}$)
\State Compute thresholds $\{u_j^{\rm ch}(t)\},\{\ell_i^{\rm dis}(t)\}$ via \eqref{eq_ex_1} and \eqref{eq_ex_2} $\forall \, (t, \text{history})$ reachable in $\mathcal{D}$
\State $\mathcal{D}^{\rm feas}\gets\varnothing$; $\mathcal{Z}\gets 0$; $\tau^{\star}\gets \varnothing$
\For{each sequence $d\in\mathcal{D}$}
  \State Compute $e_t(d)=e_0+P\sum_{\tau\le t}d_\tau$ $\forall \,t\in\mathcal{T}$
  \If{$e_t(d)\in [E_{\rm min},E_{\rm max}]$ $\forall t\in\mathcal{T}$}
    \State Compute $P^{\dagger}(d)$ via \eqref{prob_raw}, using $\alpha_t,\beta_t,\gamma_t$ from \eqref{probs}
    \State $\mathcal{D}^{\rm feas}\gets\mathcal{D}^{\rm feas}\cup\{d\}$, and $\mathcal{Z}\gets\mathcal{Z}+P^{\dagger}(d)$
  \EndIf
\EndFor
\For{each $d\in\mathcal{D}^{\rm feas}$}
  \State $\mathbb{P}_{\boldsymbol{\lambda}}\{d\mid\mathcal{D}^{\rm feas}\}\gets \tfrac{P^{\dagger}(d)}{\mathcal{Z}}$
  \State $\mathrm{dist}_T[e_T(d)]\mathrel{+}=\mathbb{P}_{\boldsymbol{\lambda}}\{d\mid\mathcal{D}^{\rm feas}\}$
\EndFor
\For{$t\in\mathcal{T}$}
  \State Repeat the renormalization on $\{t\!+\!1,\dots,T\}$ to obtain $\mathrm{dist}_t[e]$ and $q_t(e)$ for each feasible $e$
  \If{$q_t(e_t)\!\geq\!(1\!-\!\epsilon)$\textbf{and} $\tau^{\star} = \varnothing$}
    \State $\tau^{\star}\gets t$
  \EndIf
\EndFor
\State \textbf{Return:} $\mathrm{dist}_T[\cdot]$, $\tau^{\star}$, $\mathbb{P}_{\boldsymbol{\lambda}}\{e_T\in\mathcal{E}\}$
\end{algorithmic}
\end{algorithm}

\section{Conformalized quantile regression for battery energy arbitrage} \label{section_CQR}
To compare the probability estimates from Section \ref{subsection_threshold}, we implement \emph{conformalized quantile regression} (CQR). We use CQR as a distribution-free method to retrospectively validate the SoC distribution obtained under the threshold policy as in \cite{romano2019conformalized, yeh2024end}. Although CQR could also be used as a forecasting tool, this application is left for future work.

For each day $n$, let $\lambda_n$ denote the input feature vector and $e_{T,n}$ the corresponding terminal SoC obtained under the threshold policy in Sec. \ref{subsection_threshold}. The feature vector includes the realized daily price trajectory, the initial SoC, the selected pre-policy budget $(k_{\rm ch},k_{\rm dis})$, and summary statistics of the price trajectory. We split the historical data into training, calibration, and test sets, denoted by $\mathcal{D}^{\rm train}$, $\mathcal{D}^{\rm calib}$, and $\mathcal{D}^{\rm test}$, respectively. Given the test samples $\lambda^{\rm test}\!=\!\{\lambda_n^{\rm test}\}^{n\in\mathcal{D}^{\rm test}}$, CQR builds a prediction set $\mathcal{C}(\lambda^{\rm test})$ for the corresponding terminal SoC.

The goal is to guarantee marginal coverage at level $(1\!-\!\epsilon)$ across sampled energy price sequences. Accordingly, the probability estimates and the corresponding prediction set can be expressed as follows:
\begin{align*}
    \mathbb{P}_{\boldsymbol{\lambda}}\!\left\{e^{\rm test}_{T} \in \mathcal{C}(\lambda^{\rm test})\right\} &\ge (1\!-\!\epsilon),\\
    \mathcal{C}(\lambda^{\rm test}) &= \{e:\; S(\lambda^{\rm test},e)\le \hat{\Delta}\},
\end{align*}
where $S(\cdot)$ is a nonconformity score and $\hat{\Delta}$ is an empirical quantile computed using the calibration data. In particular, the set $\mathcal{C}(\cdot)$ can be taken as a box (interval) set, so that $\mathcal{C}(\lambda) = [L(\lambda),U(\lambda)]$, which can be interpreted as a sample-based approximation built from historical pairs $(\lambda, e_T)$. 

We first fit a quantile regression model for the conditional distribution $\{e_{T,n}|\lambda_n\}_{n=1}^{N}$, producing a lower conditional quantile $\hat{q}_{\rm low}$ and an upper quantile $\hat{q}_{\rm high}$. On $\mathcal{D}_{\rm train}$, we train a base quantile regression model at levels $\rho_{\rm low}\!=\!\tfrac{\epsilon}{2}$ and $\rho_{\rm high}\!=\!(1\!-\!\tfrac{\epsilon}{2})$ to obtain $\hat{q}_{\rm low}(x)\!\approx\! Q_{e_T|\lambda=x}(\rho_{\rm low})$ and $\hat{q}_{\rm high}(x)\!\approx\! Q_{e_T|\lambda=x}(\rho_{\rm high})$ via pinball loss, i.e., $L_{\theta}(y, \hat{y})\!=\!\max(\theta(y\! -\!\hat{y}), (\theta\!-\! 1)(y\! -\! \hat{y}))$ where $\theta$ is the target quantile.

For each $n \in \mathcal{D}_{\rm calib}$, we evaluate the realized price sequence under the threshold policy to obtain $e_{T,n}$ and evaluate the predicted quantiles $\hat{q}_{\rm low}(\lambda_n)$ and $\hat{q}_{\rm high}(\lambda_n)$ to compute the nonconformity score $S_n = \max [\,\hat{q}_{\rm low}(\lambda_n)\!-\!e_{T,n}, e_{T,n}\!-\!\hat{q}_{\rm high}(\lambda_n)]$,
where $S_n$ is positive exactly when $e_{T,n}$ falls outside the predicted interval $[\hat{q}_{\rm low}(\lambda_n), \hat{q}_{\rm high}(\lambda_n)]$, and negative otherwise.

We define $n_{\rm calib}\!=\!|\mathcal{D}_{\rm calib}|$ and $\gamma\!=\! \tfrac{\lceil (n_{\rm calib}\!+\!1)(1\!-\!\epsilon) \rceil}{n_{\rm calib}}$, and take the empirical $\gamma$-quantile of $\{S_n\}_{n\in \mathcal{D}_{\rm calib}}$ and denote it by $\hat{\Delta}$. This value adjusts the model interval to correct for miscoverage, in the sense that, if the model is accurate, $\hat{\Delta}\!\rightarrow\! 0$ and, otherwise, $\hat{\Delta}$ increases to correct mispredictions. Finally, for each test day $n\!\in\!\mathcal{D}_{\rm test}$, we compute the conformalized bounds $\Tilde{L}(\lambda_n)\!=\!\hat{q}_{\rm low}(\lambda_n) \!-\!\hat{\Delta}$ and $\Tilde{U}(\lambda_n)\!=\!\hat{q}_{\rm high}(\lambda_n)\!+\!\hat{\Delta}$, and define the final conformalized prediction interval for $e_{T,n}$ as follows:
\begin{equation}
    C(\lambda_n)= [\Tilde{L}(\lambda_n), \Tilde{U}(\lambda_n)],
\end{equation}
where $C(\lambda_n)$ provides a $(1\!-\!\epsilon)$ interval for the terminal SoC under the threshold policy and daily energy price sequences. In principle, if for a given day the conformal interval satisfies $C(\lambda_n)\!\subseteq\!\mathcal{E}$, then $e_T\!\in\!\mathcal{E}$ with probability at least $(1\!-\!\epsilon)$ with respect to the CQR marginal coverage guarantee.

\section{Case Study}
In this section, we present computational experiments that evaluate the proposed framework from Sec. \ref{sec:formulation} using real-world PJM data. We use the probability-redistribution procedure in Section \ref{subsection:uncertainty} to estimate the probability of finishing within a SoC target band, and to study its sensitivity to the initial state and the time at which the battery operation begins (hereafter \emph{start time window}). We use the CQR model from Sec. \ref{section_CQR} to build prediction intervals for the terminal SoC and to assess out-of-sample coverage and target band certificates.

We use six years of real-world PJM data, $\{\lambda_i\}_{i=1}^{N}$, similarly to \cite{donti2017task}, with hourly energy prices as inputs to the arbitrage policy and CQR model. The operating horizon is $T\!=\!24$ hours, with a single target time at the terminal hour, i.e., $\mathcal{T}^{\rm target}\!=\!\{24\}$. The battery operates within $E_{\rm min}\!=\!0$ and $E_{\rm max}\!=\!10$ MWh, with a maximum rate $P\!=\!2$ MW.

\subsection{Dynamic vs. static thresholds} 
\begin{figure}[ht]
    \centering
    \includegraphics[width=\linewidth]{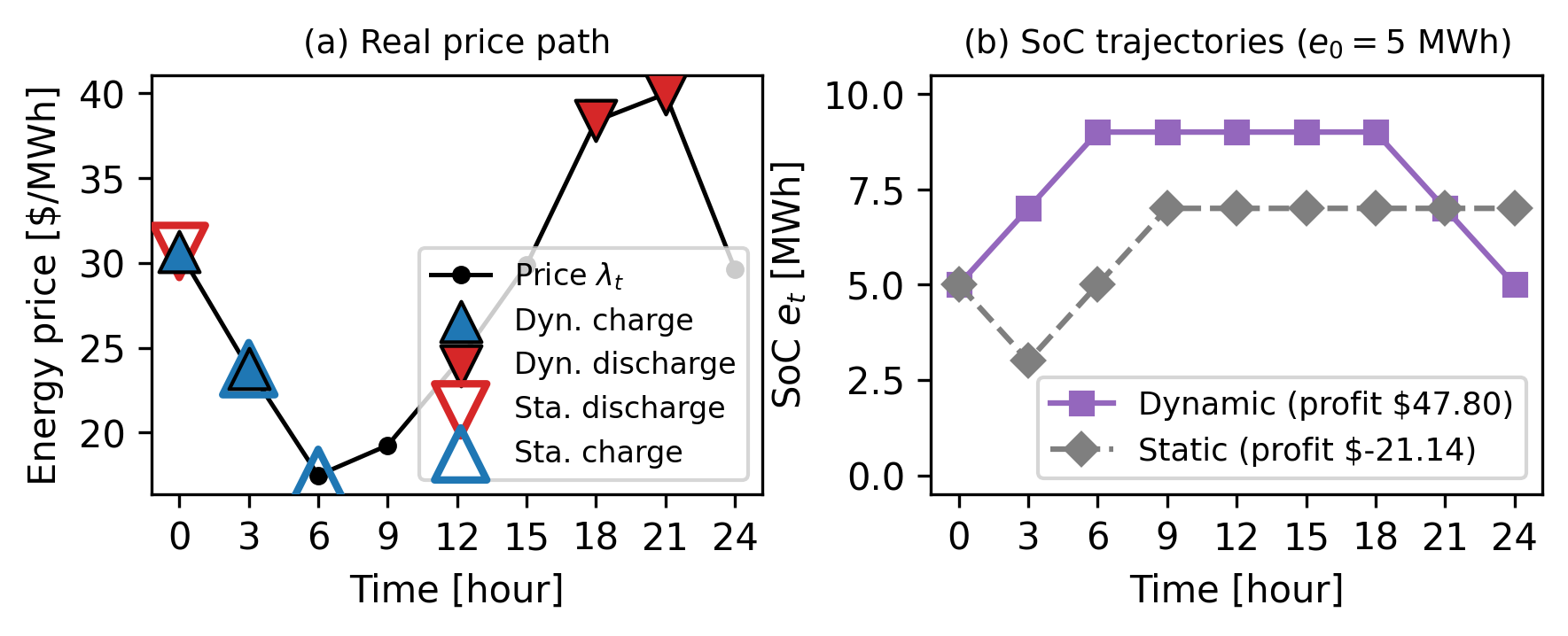}
    \caption{Dynamic vs. static threshold comparison on a real PJM price day where the two schemes' first actions diverge. (a) Realized price path with each scheme's triggered actions. (b) Resulting SoC trajectories and realized profit.}
    \label{fig:dynstatic_demo}
\end{figure}

We compare the static thresholds policy, whose thresholds are computed once from the maximum and minimum price as in \eqref{threshold_u} and \eqref{threshold_l}, against the dynamic thresholds in \eqref{eq_ex_1} and \eqref{eq_ex_2}, which by contrast are computed from the \emph{remaining} price bounds $z^{\rm min}(t), z^{\rm max}(t)$ at each $t$. Fig. \ref{fig:dynstatic_demo} uses the real PJM price data to illustrate why this time-dependence matters. The static threshold discharges at $t\!=\!0$ ($\lambda_0\!=\!\$30.68$/MWh) because that price is high relative to the minimum point, then charges twice at $t\!=\!3$ and $6$, netting a \emph{loss} of $\$21.14$ over the day. The dynamic threshold, at $t\!=\!0$, recognizes $\$30.68$ as cheap relative to the much higher prices still to come that afternoon ($\$38$-$40$/MWh) and charges instead, reserving its discharge budget for that peak; it realizes a profit of $\$47.80$. Hence, the difference of $\$68.94$ is attributed to the choice of the threshold rule. Thus, static thresholds can misjudge whether an early price is relatively cheap or expensive because they are anchored to extremes rather than to what remains achievable in the rest of the horizon.

Table \ref{table_dynstatic} reports this comparison for the full range $e_0\!\in\!\{1,\ldots,9\}$ MWh, aggregating all profitable pre-policies using their expected profits as weights. The dynamic thresholds achieve higher expected profit than static thresholds at every initial SoC, with the improvement ranging from $26.3\%$ at $e_0\!=\!1$ MWh down to $6.5\%$ at $e_0\!=\!9$ MWh, while the SoC target band probability remains essentially unchanged between the two cases at every $e_0$.
\begin{table}[ht]
  \centering
  \begin{tabular}{cccccc}
    \toprule
    $e_0$ & \multicolumn{2}{c}{Expected profit} & \multirow{2}{*}{Improv.} & \multicolumn{2}{c}{$\mathbb{P}_{\boldsymbol{\lambda}}\{e_T\!\in\![3,8]\}$} \\
    \cmidrule(lr){2-3} \cmidrule(lr){5-6}
    [MWh] & Dynamic & Static & & Dynamic & Static \\
    \midrule
    1 & 79.1 & 62.7 & $+26.3\%$ & 0.000 & 0.000 \\
    3 & 131.4 & 120.9 & $+8.6\%$ & 0.335 & 0.324 \\
    5 & 172.3 & 160.5 & $+7.4\%$ & 0.560 & 0.556 \\
    7 & 210.4 & 196.6 & $+7.0\%$ & 0.698 & 0.699 \\
    9 & 239.4 & 224.7 & $+6.5\%$ & 0.776 & 0.769 \\
    \bottomrule
  \end{tabular}
  \caption{Dynamic vs. static threshold comparison, representative $e_0$: the dynamic thresholds are more profitable at every $e_0$ without sacrificing SoC target band reliability. Band probabilities here are profit-weighted across profitable pre-policies only.}
  \label{table_dynstatic}
\end{table}
The dynamic thresholds are thus more profitable without sacrificing reliability and remain robust across a range of initial SoC values.

Since Table \ref{table_dynstatic} reports exact expected values under the assumed price distribution rather than a simulation over sampled scenarios, we verify that the same conclusion holds empirically. To this end, we re-evaluate both the static and dynamic thresholds on the same historical PJM days. Table \ref{table_dynstatic_realdays} reports the comparison of profit on a day-by-day basis.

\begin{table}[ht]
  \centering
  \begin{tabular}{ccccc}
    \toprule
    $e_0$ & \multicolumn{2}{c}{Mean realized profit [\$]} & \multirow{2}{*}{Improv.} & Days dynamic \\
    \cmidrule(lr){2-3}
    [MWh] & Dynamic & Static & & wins [\%] \\
    \midrule
    1 & $17.9$ & $-0.6$ & $-$ & $77.1$ \\
    3 & $45.8$ & $37.1$ & $+23.5\%$ & $42.3$ \\
    5 & $69.3$ & $52.9$ & $+31.2\%$ & $61.3$ \\
    7 & $84.5$ & $60.9$ & $+38.7\%$ & $70.2$ \\
    9 & $91.7$ & $65.4$ & $+40.1\%$ & $71.5$ \\
    \bottomrule
  \end{tabular}
  \caption{Empirical dynamic vs. static threshold comparison on real historical PJM days.}
  \label{table_dynstatic_realdays}
\end{table}

Table \ref{table_dynstatic_realdays} shows that the dynamic policy achieves a higher mean realized profit for every $e_0$. Thus, gains range from $23.5\%$ to $40.1\%$ for $e_0\! \in \!\{2,\ldots,9\}$, while at $e_0\!=\!1$ MWh the result shifts from an average loss under static thresholds ($-0.6$) to a positive average profit under dynamic thresholds ($+17.9$). This advantage holds in expectation rather than on every individual day. The fraction of days on which the dynamic policy outperforms the static policy ranges from $42\%$ to $77\%$ across $e_0$ values, indicating that the larger gains on winning days outweigh the losses on the remaining days.

As a baseline, we also compare against a passive operator that never trades in real time and holds its initial SoC throughout, i.e., the BO takes no corrective action, resulting in $e_T\!=\!e_0$ and zero arbitrage profit. At $e_0\!=\!5$ MWh, the threshold policy instead achieves an expected profit of $\$172.3$. The passive operator case reaches $\mathcal{E}$ with probability $1$ only because the initial SoC lies within both target bands. Therefore, this high-reliability result is driven by the initial condition. Conversely, for any $e_0\notin\mathcal{E}$, the passive operator reaches $\mathcal{E}$ with probability $0$ by construction, whereas the threshold policy retains a positive probability.

\subsection{Depletion mitigation: an illustrative example}
\begin{figure}[ht]
    \centering
    \includegraphics[width=1.05\linewidth]{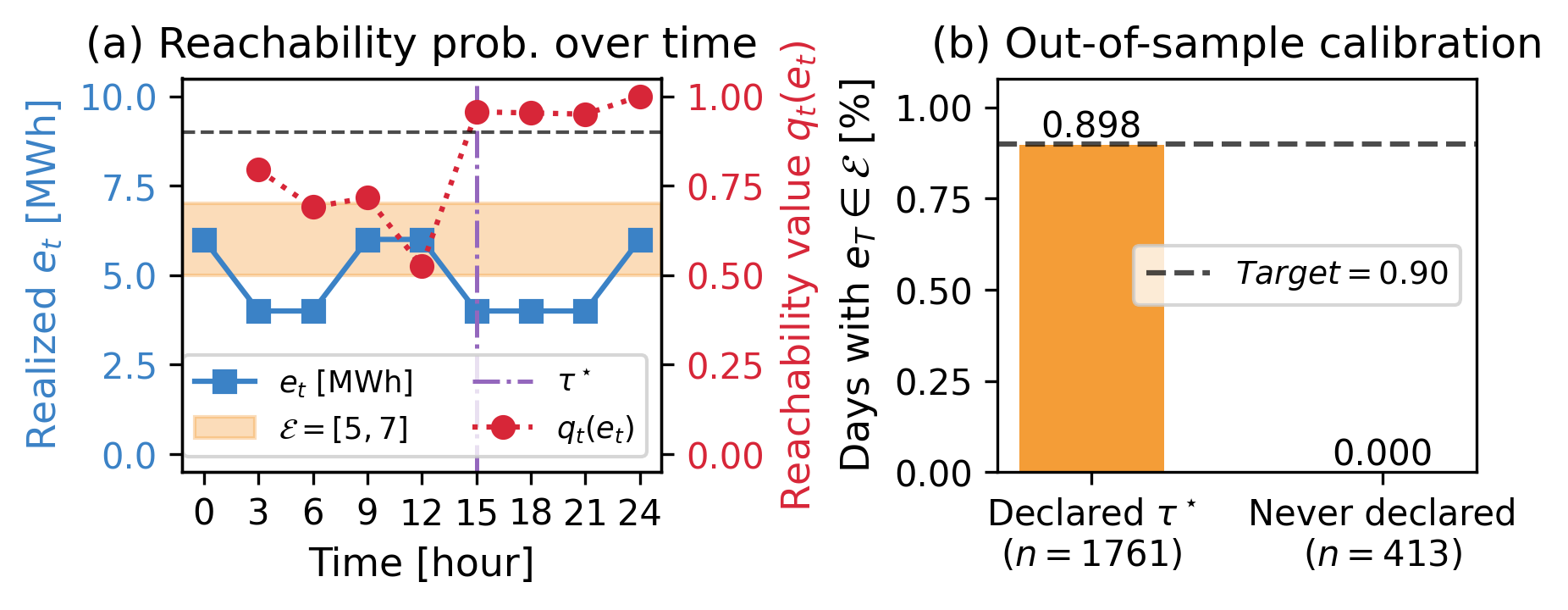}
    \caption{Stopping time depletion-risk signal on real PJM data. (Left) Illustrative realized SoC and reachability probability $q_t(e_t)$, which crosses $(1\!-\!\epsilon)$ at $\tau^{\star}\!=\!5$. (Right) Out-of-sample calibration that shows the empirical fraction of days landing in $\mathcal{E}$, conditioned on whether $\tau^{\star}$ is reached within the horizon.}
    \label{fig:depletion_demo}
\end{figure}

We consider the battery with $\mathcal{E}=[5,7]$ MWh and $e_0\!=\!6$ MWh, and validate the stopping time signal on $2174$ independent, out-of-sample days. Here, $\tau^{\star}$ is declared within the horizon on $1761$ of these days. Among them, the empirical fraction that actually lands in $\mathcal{E}$ is $89.8\%$, roughly matching the target confidence level $(1\!-\!\epsilon)\!=\!90\%$. Consistently with this assumption, on the remaining $413$ days, $q_t(e_t)$ never crosses $(1\!-\!\epsilon)$ within the horizon and none end in $\mathcal{E}$. Fig. \ref{fig:depletion_demo} shows the depletion risk identified through this calibration check. The clear separation in Fig. \ref{fig:depletion_demo} (right) between the days on which $\tau^{\star}$ is reached and those on which it is not is a well-calibrated indicator of depletion risk. Whenever the reachability probability falls below $(1\!-\!\epsilon)$, the real-data test indicates that the terminal SoC is likely to end outside $\mathcal{E}$, providing an early warning signal needed for the SO to act before depletion becomes unavoidable.

\subsection{Impact of Pruning on Reachability} We assess the impact of the proposed Algorithm \ref{alg:sop} by comparing the \emph{raw} capacity-unconstrained probability in \eqref{prob_target_band} against the feasibility-conditioned $\mathrm{dist}_T[\cdot]$ obtained from \eqref{eq_renorm}. Table \ref{table_ablation} reports the comparison over the full range $e_0\!\in\!\{1,\ldots,9\}$ MWh. The bias is negative (i.e., the raw estimate understates reachability) for every $e_0$ and both target bands, with an average of $-22.8\%$ and a range from $8.1\%$ to $41.9\%$. For instance, at $e_0\!=\!5$ MWh, disabling the pruning algorithm understates $\mathbb{P}_{\boldsymbol{\lambda}}\{e_T\!\in\!\mathcal{E}\}$ by $0.217$ for $\mathcal{E}=[5,7]$ and by $0.345$ for $\mathcal{E}=[3,8]$. These results empirically confirm that pruning materially changes, rather than merely refines, the reachability estimate.
\begin{table}[ht]
  \centering
  \begin{tabular}{ccccccc}
    \toprule
    \multirow{2}{*}{$e_0$} & \multicolumn{2}{c}{Bias (\%)} & \multicolumn{2}{c}{$\mathbb{P}_{\boldsymbol{\lambda}}\{e_T\!\in\![5,7]\}$} & \multicolumn{2}{c}{$\mathbb{P}_{\boldsymbol{\lambda}}\{e_T\!\in\![3,8]\}$} \\
    \cmidrule(lr){2-3}\cmidrule(lr){4-5}\cmidrule(lr){6-7}
    [MWh] & $\mathcal{E}=[5,7]$ & $[3,8]$ & \textit{Raw} & \textit{Pruned} & \textit{Raw} & \textit{Pruned} \\
    \midrule
    1 & $-27.3$ & $-27.2$ & $0.204$ & $0.478$ & $0.312$ & $0.584$\\
    3 & $-22.9$ & $-32.3$ & $0.213$ & $0.442$ & $0.311$ & $0.633$\\
    5 & $-21.7$ & $-34.5$ & $0.205$ & $0.422$ & $0.306$ & $0.651$\\
    7 & $-17.7$ & $-33.9$ & $0.198$ & $0.376$ & $0.281$ & $0.620$\\
    9 & $-16.3$ & $-41.9$ & $0.183$ & $0.346$ & $0.255$ & $0.673$\\
    \bottomrule
  \end{tabular}
  \caption{Pruning ablation by $e_0$. Bias of the raw SoC target band probability estimate relative to the pruned, alongside the underlying chance-constraint probabilities $\mathbb{P}\{e_T\!\in\!\mathcal{E}\}$ for both target bands.}
  \label{table_ablation}
\end{table}

\subsection{Initial SoC and start time window sensitivity}
We vary $e_{t=0}\!\in\!\{1,\ldots,9\}$ MWh, and consider two terminal SoC target bands, $\mathcal{E}\!=\![5,7]$ and $[3,8]$ MWh, to show the sensitivity of the resulting attainment probabilities to $e_0$ and $\mathcal{E}$. We evaluate multiple different start time windows (e.g., running the algorithm from $t\!=\!3$ instead of $t\!=\!0$), estimating the probability of satisfying the SoC target band, and quantifying the associated profit outcomes under the threshold policy.
\begin{figure}[ht]
  \centering
  \begin{subfigure}[t]{0.48\linewidth}
    \centering
    \includegraphics[width=\textwidth, trim={0 2.3mm 0 0mm},clip]{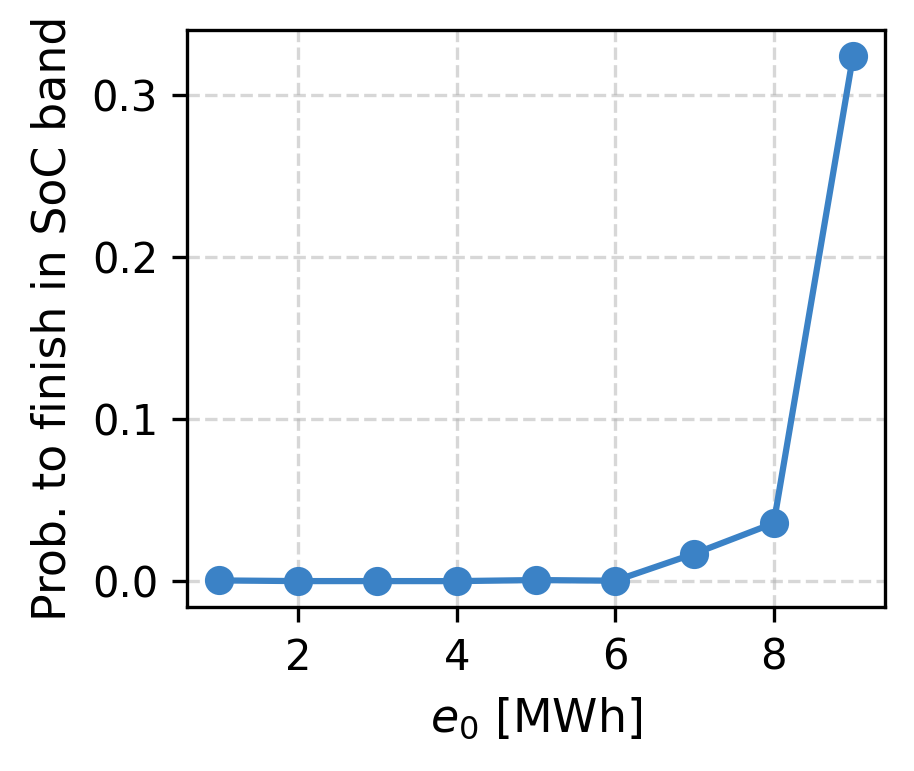}
    \caption{Attain. prob. vs. $e_0$ at $t=0$}
    \label{fig:multi_band_3}
  \end{subfigure}
  \hfill
  \begin{subfigure}[t]{0.48\linewidth}
    \centering
    \includegraphics[width=1.05\textwidth, trim={0 2.3mm 0 6.5mm},clip]{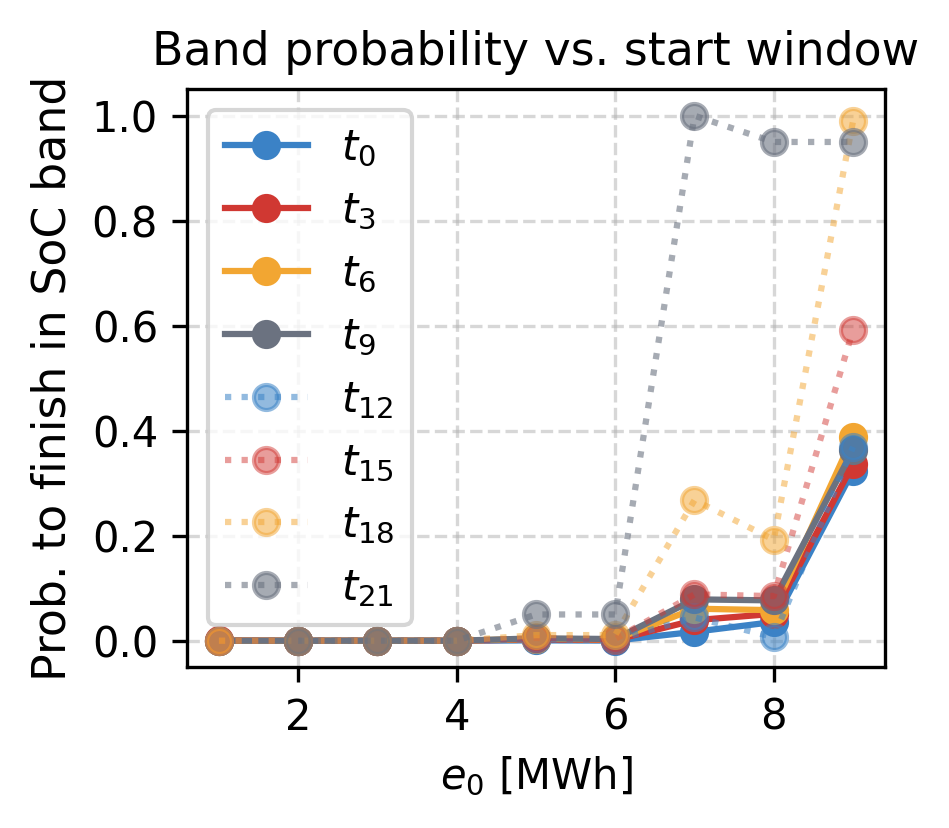}
    \caption{Attain. prob. vs. start time}
    \label{fig:multi_band_4}
  \end{subfigure}
  \\[1mm]
  \begin{subfigure}[t]{\linewidth}
    \centering
    \includegraphics[width=\textwidth, trim={0 2.5mm 0 8mm},clip]{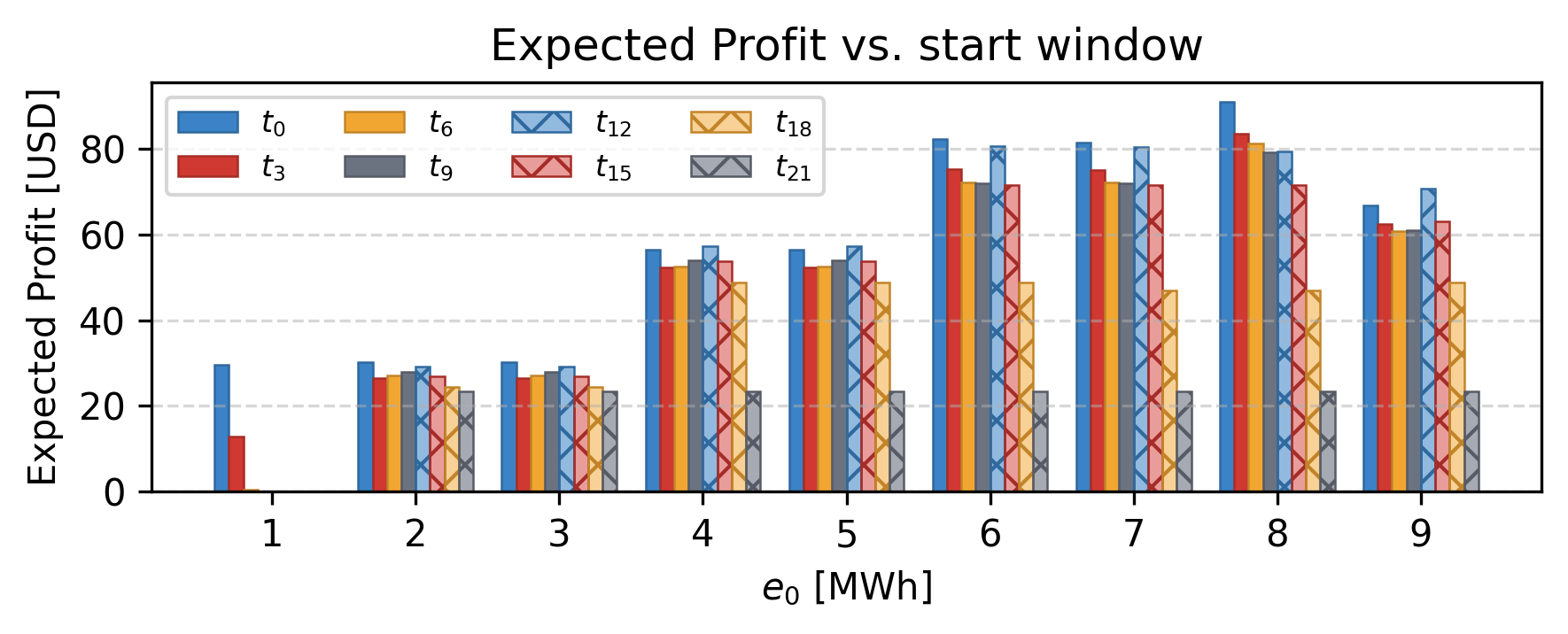}
    \caption{Profit vs.\ start time}
    \label{fig:multi_band_5}
  \end{subfigure}
  \caption{Sensitivity of SoC target band $\mathcal{E}=[5,7]$ MWh attainment and profit to initial SoC $e_0$ and start time window. (a) Attainment probability vs. $e_0$. (b) Attainment probability vs. start time. (c) Accumulated profit vs. start time window.}
  \label{fig:multi_band_345}
\end{figure}

Fig. \ref{fig:multi_band_345} shows how the probability of attaining the tighter SoC target band $\mathcal{E}=[5,7]$ MWh, and the resulting profit, depend on $e_0$ and the start time. Fig. \ref{fig:multi_band_3} shows that this probability is highly sensitive to $e_0$, i.e., when $e_0$ is low, the probability of finishing within $\mathcal{E}$ is close to zero, and it increases only when $e_0$ is already near the upper range of $\mathcal{E}$. This behavior is driven by the limited rate $P$, the finite number of available actions over the horizon, and the requirement that the battery follows a policy with positive expected profit.
The same objective explains the counterintuitive charging behavior at low $e_0$. Since \eqref{P1:objective_function} rewards only profitable buy/sell spreads within the horizon, the pre-policy tends to choose similar numbers of charging and discharging actions ($k_{\rm ch}\!\approx\!k_{\rm dis}$). As a result, it places no value on ending with a higher SoC unless that energy can later be sold at a profit. Consequently, when $e_0$ starts far below $\mathcal{E}$, only a small set of feasible trajectories can reach $\mathcal{E}$. Fig. \ref{fig:multi_band_4} shows the same probability varying the start time. Intuitively, initiating the policy later reduces the remaining flexibility to correct the SoC trajectory and limits the available arbitrage opportunities. When $e_0$ is already above $\mathcal{E}$, starting later may reduce the number of trajectories that drift the terminal SoC outside the band. Thus, the probability to finish within the SoC target band depends on the state (through $e_0$) and time (through the start time and realized price sequence), which makes the stopping time $\tau^{\star}$ difficult to satisfy across all values of $e_0$. Fig. \ref{fig:multi_band_5} shows the corresponding accumulated profit across start time windows and initial SoC values. The profit outcomes vary across start time windows, since the intraday price differs depending on when the policy is initiated, and conservative actions taken to preserve feasibility can reduce the exploitation of arbitrage opportunities. Together with Fig. \ref{fig:multi_band_4}, these results highlight a SoC target reachability and profit trade-off. Dispatching the battery to increase the probability of finishing within $\mathcal{E}$ can reduce the attainable arbitrage profit under the same realized price sequence. 

\begin{table}[ht]
  \centering
  \begin{tabular}{cccccc}
    \toprule
    \multirow{2}{*}{Initial SoC} & \multirow{2}{*}{SoC band} & \multicolumn{4}{c}{Start time hour} \\
    \cmidrule(lr){3-6}
    [MWh] & [MWh] & 0 & 6 & 12 & 18 \\
    \midrule
    \multirow{2}{*}{1} & [5,7] & 46.66 & 43.63 & 37.14 & 20.00 \\
    & [3,8] & 73.17 & 72.97 & 71.43 & 60.00 \\
    \multirow{2}{*}{5} & [5,7] & 50.01 & 50.10 & 50.72 & 55.56 \\
     & [3,8] & 73.22 & 73.31 & 73.91 & 77.78 \\
    \multirow{2}{*}{9} & [5,7] & 53.29 & 55.98 & 60.00 & 60.00 \\
     & [3,8] & 73.17 & 72.97 & 71.43 & 60.00 \\
    \bottomrule
  \end{tabular}
  \caption{Percentage of feasible trajectories that finish within the SoC target band, by initial SoC, and start time.}
  \label{table_1}
\end{table}

Table \ref{table_1} shows the ratio between the number of SoC trajectories that lead to $\mathcal{E}$ and the total number of feasible trajectories across different initial SoC values and start time windows (without considering the battery policy). The percentage of feasible trajectories reaching a tight target band $\mathcal{E}\!=\![5,7]$ MWh varies significantly with $e_0$. For $e_0\! \geq\! 3$ MWh, this percentage generally increases with $e_0$ and the start time, whereas for $e_0\! \leq\! 3$ it tends to decrease as the start time shifts later. When the band is wider, $\mathcal{E} = [3,8]$ MWh, the overall percentage increases, but for initial SoC values outside the target band, the percentage of trajectories tends to decrease with $e_0$ and with the start time.
\begin{figure}[ht]
  \centering
  \begin{subfigure}[t]{0.48\linewidth}
    \centering
    \includegraphics[width=1.05\textwidth, trim={0 1.5mm 0 6.5mm},clip]{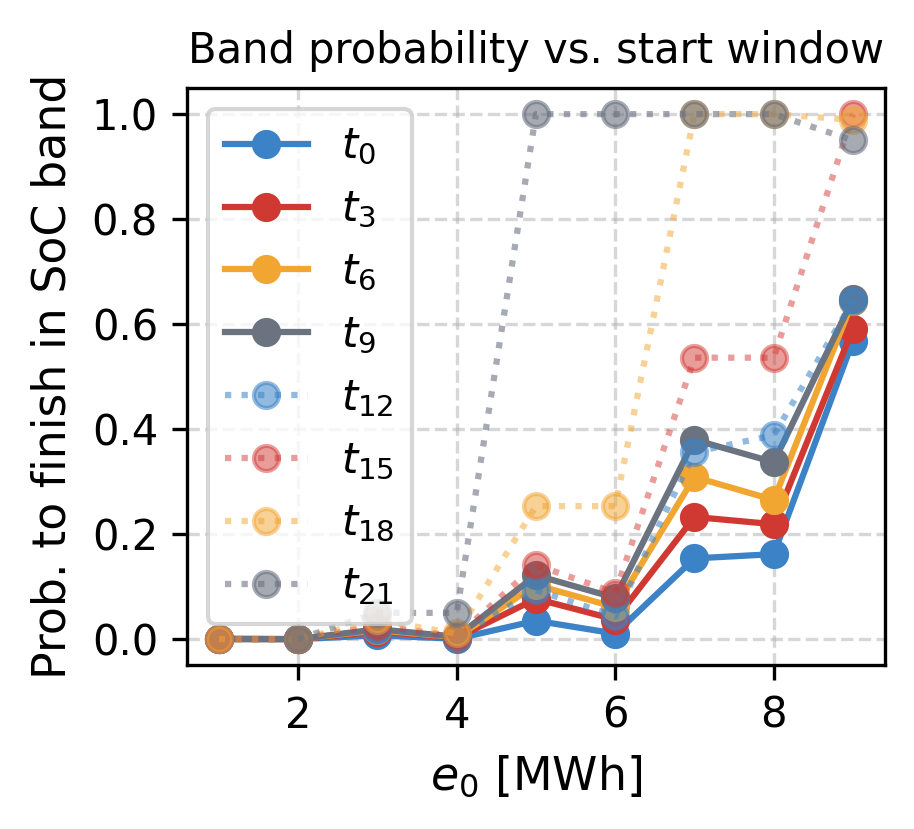}
    \caption{vs.\ start time}
    \label{fig:multi_band_6}
  \end{subfigure}
  \hfill
  \begin{subfigure}[t]{0.48\linewidth}
    \centering
    \includegraphics[width=1.07\textwidth, trim={0 -2.3mm 0 6.5mm},clip]{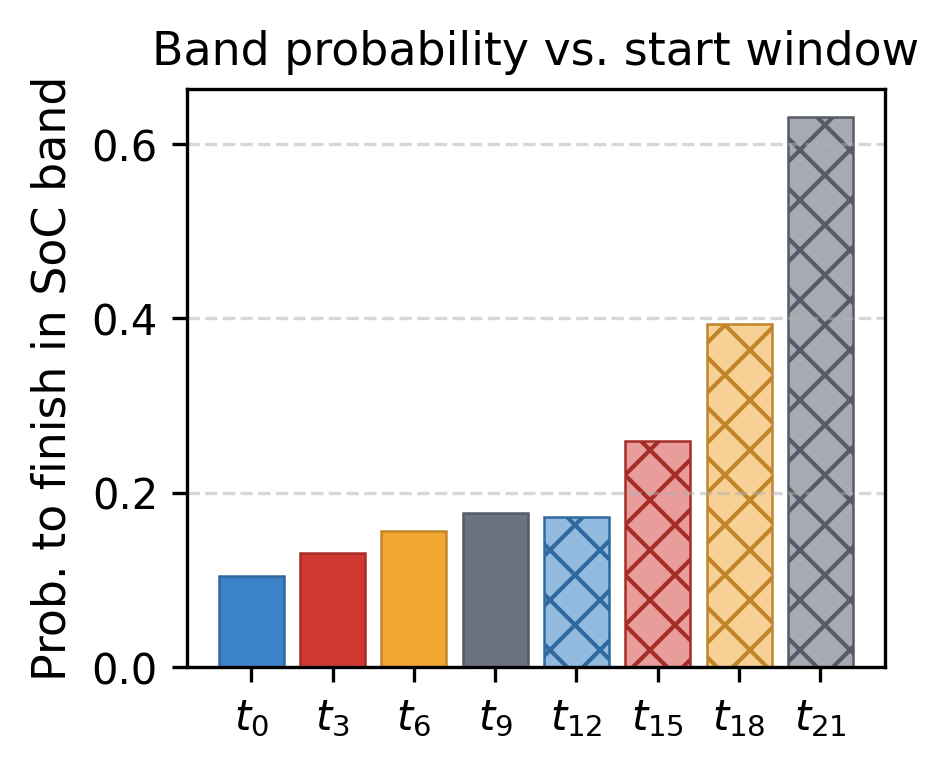}
    \caption{averaged across $e_0$}
    \label{fig:multi_band_8}
  \end{subfigure}
  \caption{Sensitivity of the wider SoC target band $\mathcal{E}=[3,8]$ MWh attainment probability to start time and initial SoC $e_0$. (a) Attainment probability vs.\ start time, across $e_0$. (b) Average attainment probability across $e_0$, vs.\ start time.}
  \label{fig:multi_band_68}
\end{figure}

Fig. \ref{fig:multi_band_68} shows the same analysis for the wider band $\mathcal{E}=[3,8]$ MWh. In particular, Fig. \ref{fig:multi_band_6} shows that widening $\mathcal{E}$ increases success probabilities across most initial SoC values and start time windows, as expected, but the dependence on both $e_0$ and start time remains. Probability curves are not uniform across start time windows even under this wider interval, indicating persistent time-of-day effects, and the stopping time $\tau^{\star}$ retains a stronger dependence on $e_0$. For example, $\tau^{\star}\!=\!21$ hr guarantees reachability for $e_0\!\geq\!5$ MWh but the associated probability falls below $0.2$ for $e_0\!\leq\!5$ MWh, and the same pattern holds at $\tau^{\star}\!=\!15$ hr with the threshold shifted to $e_0\!\geq\!7$ MWh. Fig. \ref{fig:multi_band_8} aggregates the start windows by averaging the attainment probability across initial SoC values. The results show an overall increasing trend for later start windows, suggesting that, under the proposed battery policy, a state-independent stopping time $\tau^{\star}$ cannot provide reliability guarantees above approximately $60\%$.

Table \ref{table_violation} reports the empirical fraction of realized PJM days for which $e_T\!\notin\!\mathcal{E}$. Violation rates are highest for low $e_0$, reaching $0.992$ for $\mathcal{E}=[5,7]$ at $e_0=1$ and $0.998$ at $e_0=2$. They decrease as $e_0$ approaches the target band but remain substantial, with a minimum observed rate of $0.394$ at $e_0=7$ for $\mathcal{E}=[3,8]$. These results provide a realization-based counterpart to the theoretical $(1\!-\!\epsilon)=0.9$ reachability target under a state-independent policy. Table \ref{table_violation} also reports the theoretical attainment probability computed from the assumed price distribution and the pruning and renormalization step of Sec. \ref{subsection:pruning}, for a direct, explicit verification of the reachability claim against realized outcomes.
\begin{table}[ht]
\centering
\begin{tabular}{ccccccc}
\toprule
& \multicolumn{2}{c}{$\mathbb{P}_{\boldsymbol{\lambda}}\{e_T\!\notin\!\mathcal{E}\}$} & \multicolumn{2}{c}{$\mathbb{P}_{\boldsymbol{\lambda}}\{e_T\!\in\!\mathcal{E}\}$} & \multicolumn{2}{c}{Theoretical $\mathbb{P}\{e_T\!\in\!\mathcal{E}\}$}\\
\cmidrule(lr){2-3} \cmidrule(lr){4-5} \cmidrule(lr){6-7}
$e_0$ [MWh] & $[5,7]$ & $[3,8]$ & $[5,7]$ & $[3,8]$ & $[5,7]$ & $[3,8]$\\
\midrule
1 & 0.992 & 0.907 & 0.008 & 0.093 & 0.478 & 0.584 \\
3 & 0.967 & 0.770 & 0.033 & 0.230 & 0.442 & 0.633 \\
5 & 0.805 & 0.560 & 0.195 & 0.440 & 0.423 & 0.651 \\
7 & 0.567 & 0.394 & 0.433 & 0.606 & 0.376 & 0.620 \\
9 & 0.557 & 0.403 & 0.443 & 0.597 & 0.346 & 0.673 \\
\bottomrule
\end{tabular}
\caption{Out-of-sample violation and attainment rates ($\%$ of realized days) for SoC target bands and initial SoC $e_0$, alongside the theoretical attainment probability.}
\label{table_violation}
\end{table}

\begin{figure}[ht]
  \centering
  \begin{subfigure}[t]{0.48\linewidth}
        \centering
        \includegraphics[width=1.02\linewidth, trim={0 2mm 0 0mm},clip]{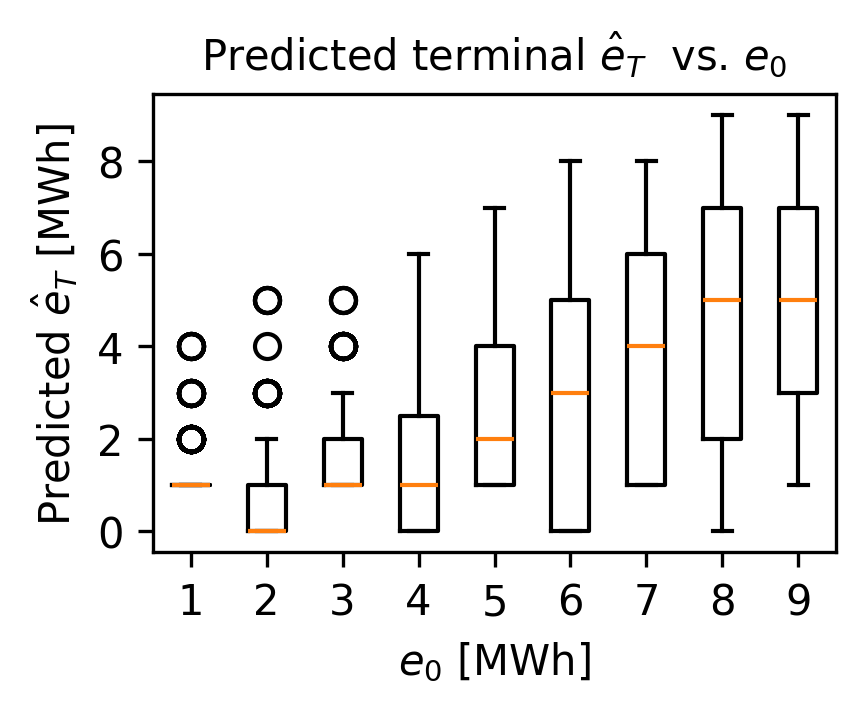}
        \label{fig:multi_band_9}
  \end{subfigure}
  \begin{subfigure}[t]{0.48\linewidth}
        \centering
        \includegraphics[width=1.05\linewidth, trim={0 2mm 0 0mm},clip]{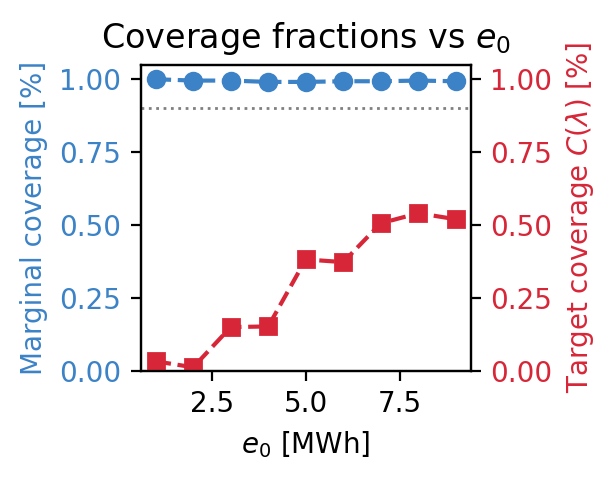}
        \label{fig:multi_band_10}
  \end{subfigure}
  \vspace{-3mm}
  \caption{(Left) Distribution of predicted terminal SoC across test data for each initial SoC and $(1\!-\!\epsilon)\! =\! 0.9$. (Right) Coverage versus the initial SoC. The blue curve shows the empirical test-set coverage and the red curve shows the SoC target band coverage between different $e_0$ values.}
  \label{fig:multi_band_910}
\end{figure}

The theoretical value is given by $\mathrm{dist}_T[\cdot]$ (described in Sec. \ref{subsection:pruning}) which treats actions across time as independent draws from the calibrated time step price distributions. In the PJM data, however, prices within a day exhibit persistent temporal patterns. A day with sufficiently low early prices to trigger several charging actions is therefore unlikely to also exhibit the later high prices needed to trigger enough discharging actions to reach a distant target band. By not capturing this intraday dependence, the theoretical estimate assigns excessive probability to trajectories with large net SoC changes. As a result, the discrepancy is largest when $e_0$ is far from $\mathcal{E}$ and narrows as $e_0$ approaches the target band.

Fig. \ref{fig:multi_band_910} (left) shows the distribution of the predicted terminal SoC $\hat{e}_T= \tfrac{(\hat{q}_{\rm low}(\lambda)+\hat{q}_{\rm high}(\lambda))}{2}$ from the CQR model across test days for each $e_0$, at price coverage level $(1\!-\!\epsilon)$ with $\mathcal{E}\!=\![5,7]$. As $e_0$ increases, the predicted terminal SoC tends to increase, and its dispersion also grows. This indicates that $e_T$, under the threshold policy, is mainly driven by $e_0$.

Fig. \ref{fig:multi_band_910} (right) shows how the out-of-sample guarantees vary with $e_0$. The empirical marginal coverage stays above the nominal $(1\!-\!\epsilon)$, remaining close to one for most $e_0$ (a conservative over-coverage induced by the discrete support of $e_{T}$), indicating that the conformalized intervals preserve the desired marginal coverage for the terminal SoC induced by the threshold policy. In contrast, the SoC target band coverage, the fraction of days for which $C(\lambda)\subseteq \mathcal{E}$, is small but non-zero for low $e_0$ ($0.03$ at $e_0=1$, $0.01$ at $e_0=2$), rises through the middle of the range, and reaches its highest value of $0.54$ at $e_0=8$ MWh before easing slightly to $0.52$ at $e_0=9$ MWh. Since the CQR does not enforce $C(\lambda)\subseteq \mathcal{E}$, some coverage gap relative to the raw attainment rate remains expected by construction, i.e., there is no guarantee that the resulting interval lies entirely inside $\mathcal{E}$. Thus, the chance constraint in \eqref{P1:chance_constraint} is rarely satisfied for low $e_0$ and improves for larger $e_0$ values, because the policy is more likely to meet the fixed band $\mathcal{E}$, when the initial state is nearer to the prescribed band.

\section{Conclusion and Future Work}
This paper presents a framework for battery energy arbitrage under uncertain energy prices that integrates an approximate chance-constrained terminal SoC requirement, defined by a safe SoC target band, with online threshold policies. Using a $k$-search structure, we derive explicit buy and sell thresholds, and compute the induced action and terminal SoC probabilities over the planning horizon. To obtain exact SoC distributions under operational limits, we develop a probability redistribution pruning algorithm and use it to quantify the probability of staying within the SoC target band and to estimate the stopping time required to satisfy the SoC chance constraint.

Results on historical PJM data indicate that the terminal SoC is highly sensitive to the initial SoC ($e_0$) and the operation start time. When $e_0$ starts far from the terminal SoC target band, the limited charge/discharge rate and the finite operation horizon leave few feasible trajectories that can reach the SoC target band. Similarly, initiating the policy later (i.e., a later start time) reduces the remaining flexibility to correct the SoC trajectory and limits available arbitrage opportunities, yielding inconsistent success probabilities across start time windows. While widening the terminal band increases success probabilities, the dependence on both state and time persists, making a single state-independent stopping time $\tau^\star$ difficult to satisfy across all cases. These results also highlight a reliability/profit trade-off, where increasing the probability of finishing within the SoC target band can reduce the attainable arbitrage profit under the same realized price sequence. In addition, CQR preserves the desired marginal out-of-sample coverage for the terminal SoC under the threshold policy, but does not yield prediction intervals contained within a prescribed SoC target band, which explains the observed gap between marginal and target band coverage for small $e_0$ values.

The framework relies on several simplifying assumptions that make the terminal SoC distribution tractable and admits the closed-loop policy for the price threshold analysis, which in turn limit the scope of the guarantees. In particular, efficiency losses, degradation costs, and continuous dispatch levels are not captured and are left for future work.


\bibliographystyle{ieeetr}
\bibliography{bib}

\appendices
\section{Threshold policy competitive guarantee}\label{app:proofs}
First, we present the competitive-guarantee proof for the time-dependent thresholds. Second, we outline how that guarantee can be preserved when the effective $k$ value is reduced mid-horizon by enforcing the same structural change on the benchmark. Third, we discuss how the feasibility constraints induced by the SoC capacity limits modify the set of admissible charge and discharge actions, and how this leads to feasibility-dependent predicted price sets and thresholds.
\subsection{Competitive guarantee proof for time-dependent thresholds}\label{app:proof_outline}
We present the proof for the discharge policy. The same arguments can be extended to the charge action. We consider the following time-dependent discharge threshold rule:
\begin{equation}\label{eq:app_time_dependent_thresh}
\ell_i^{\rm dis}(t)
= \tfrac{\omega}{k_{\rm dis}}
\Bigg( \sum_{\hat{t}=1}^{t-1}\sum_{n=1}^{i-1} \ell_{n}^{\rm dis \star}(\hat{t})
+ \sum_{m=1}^{k_{\rm dis}-(i-1)} \hat{z}_{m}^{\rm min}(t) \Bigg),
\end{equation}
where $\omega$ is the target competitive ratio and $\hat{z}_{m}^{\rm min}(t)$ is the $m$-th smallest predicted price among future times. Intuitively, when no prediction is available the set $\hat{z}_{m}^{\rm min}(t)=\lambda^{\rm min}$ for all $m$ and $t$. In this case, the term inside parentheses upper bounds the offline benchmark for the remaining discharge decisions. We denote this bound by the offline benchmark $\text{OPT}_{v}^{i}(t)$ that satisfies:
\begin{equation}\label{eq:app_vopt_ub}
\text{OPT}_{v}^{i}(t) \leq
\Bigg(\sum_{\hat{t}=1}^{t-1}\sum_{n=1}^{i-1} \ell_{n}^{\rm dis \star}(\hat{t})
+ \sum_{m=1}^{k_{\rm dis}-(i-1)} \hat{z}_{m}^{\rm min}(t) \Bigg).
\end{equation}
By construction, \eqref{eq:app_time_dependent_thresh} guarantees the competitive bound by enforcing the following constraint:
\begin{equation}\label{eq:app_thresh_vopt_relation}
\ell_i^{\rm dis}(t) = \tfrac{\omega}{k_{\rm dis}}\,\text{OPT}_{v}^{i}(t).
\end{equation}

This condition is precisely what is needed to establish the competitive guarantee, following the reservation-price technique of \cite{lorenz2009optimal}.

\begin{theorem}\label{thm:app_comp_discharge}
Let $\ell_{k_{\rm dis}+1}^{\rm dis}(\cdot) := \lambda^{\rm max}$. Under the threshold rule \eqref{eq:app_time_dependent_thresh}--\eqref{eq:app_thresh_vopt_relation}, for every realized price path consistent with the assumed price bounds, the discharge policy satisfies $\mathrm{ALG}_{\rm dis} \geq \tfrac{1}{\omega}\,\mathrm{OPT}_{\rm dis}$, where $\mathrm{OPT}{\rm dis}$ is the sum of the $k{\rm dis}$ highest realized prices over the horizon.
\end{theorem}

\begin{proof}
For $0!\leq\! j\!\leq k_{\rm dis}$, let $\mathcal{P}j$ denote the set of realized price paths for which exactly $j$ of the $k{\rm dis}$ discharge thresholds are triggered before the horizon ends. Fix $\sigma\!\in\!\mathcal{P}j$, let $\ell_1^{\rm dis\star},\ldots,\ell_j^{\rm dis\star}$ denote the realized trigger prices, and let $t_j$ be the time of the $j$-th trigger. Since $\sigma\!\in\!\mathcal{P}j$, no price after $t_j$ clears $\ell_{j+1}^{\rm dis}(\cdot)$, otherwise a $(j\!+\!1)$-th trigger would occur. Hence, $p{\max}(\sigma)\leq \ell_{j+1}^{\rm dis}(t_j)$. Since $\mathrm{OPT}_{\rm dis}$ selects at most $k_{\rm dis}$ prices, each bounded by $p_{\max}(\sigma)$,
\begin{equation}
\mathrm{OPT}_{\rm dis}(\sigma) \leq k_{\rm dis}\,\ell_{j+1}^{\rm dis}(t_j). \label{eq:app_opt_ub_case}
\end{equation}
In the worst case, the remaining $(k_{\rm dis}\!-\!j)$ discharges occur at the pessimistic forecast floors available at $t_j$, so
\begin{equation}
\mathrm{ALG}_{\rm dis}(\sigma) \geq \sum_{n=1}^{j}\ell_n^{\rm dis\star} + \sum_{m=1}^{k_{\rm dis}-j} \hat{z}_m^{\rm min}(t_j). \label{eq:app_alg_lb_case}
\end{equation}
By \eqref{eq:app_thresh_vopt_relation} applied at $i=j+1$, the right hand side of \eqref{eq:app_alg_lb_case} equals $\text{OPT}_v^{j+1}(t_j) = \tfrac{k_{\rm dis}}{\omega}\ell_{j+1}^{\rm dis}(t_j)$. Combining with \eqref{eq:app_opt_ub_case},
\begin{equation}
\frac{\mathrm{OPT}_{\rm dis}(\sigma)}{\mathrm{ALG}_{\rm dis}(\sigma)} \leq \frac{k_{\rm dis}\,\ell_{j+1}^{\rm dis}(t_j)}{\tfrac{k_{\rm dis}}{\omega}\ell_{j+1}^{\rm dis}(t_j)} = \omega.
\end{equation}
This holds for every $j\!\in\!\{0,\ldots,k_{\rm dis}\}$. For $j=k_{\rm dis}$, using $\ell_{k_{\rm dis}+1}^{\rm dis}=\lambda^{\rm max}$, it reduces to $\mathrm{OPT}_{\rm dis}\leq k_{\rm dis}\lambda^{\rm max}$. Since $\{\mathcal{P}_j\}_{j=0}^{k_{\rm dis}}$ partitions all admissible price paths, the bound holds for every realized path $\sigma$, proving
$\mathrm{ALG}_{\rm dis}\geq\tfrac{1}{\omega}\mathrm{OPT}_{\rm dis}$.
\end{proof}

Setting $\hat{z}_m^{\rm min}(t)\equiv\lambda^{\rm min}$ for all $m,t$ recovers the static reservation-price policy of \cite{lorenz2009optimal}. With time-varying lower bounds, the forecast sequence must remain consistent, meaning that later updates may refine but not contradict previously revealed prices. This preserves the admissibility of the pessimistic completion in \eqref{eq:app_alg_lb_case}. By exchanging $\lambda^{\rm min}$ and $\lambda^{\rm max}$ and replacing $\ell^{\rm dis}$ with $u^{\rm ch}$, the same argument gives $\mathrm{ALG}{\rm ch}\leq\omega\,\mathrm{OPT}_{\rm ch}$ for \eqref{eq_ex_1}.

\subsection{Mid-horizon reduction of $k$ for policy decision}\label{app:mid_horizon_k}
Let $G^{(k)}$ be the threshold outcome of the optimal $k$-max method, $\mathrm{OPT}^{(k)}$ be the offline optimum constrained to discharge $k$ units, and $R(\cdot)$ be the associated reward as a function of the price sequence. The standard $\omega$-competitive guarantee, where $\omega$ denotes the competitive ratio, can be expressed for any price sequence $p_{(1:T)}\! =\! \{p_t\}_{t=1}^{T}$ as follows:
\begin{equation}\label{eq:app_comp_ratio}
R(G^{(k)}, p_{(1:T)}) \geq \tfrac{1}{\omega}\,\text{OPT}^{(k)}(T).
\end{equation}
Suppose that at time $\Bar{t}-1$ the policy has activated $s$ discharge actions, and the remaining objective is reduced from $k$ to $\Bar{k}<k$ for $t\geq \Bar{t}$. Over the remaining periods $[\bar{t},T]$, the policy is required to discharge $k_{\rm rem}\!=\!\bar{k}-s$ units. Let $G^{(k_{\rm rem})}$ be the optimal $k$-search algorithm for the remaining periods, with its own reservation prices on the sub-horizon $p_{(\Bar{t}:T)}$. Under this change, the original bound in \eqref{eq:app_comp_ratio} does not apply directly if the benchmark remains $\text{OPT}^{(k)}$, because the online and offline problems no longer share the same structure after $\Bar{t}$.

To preserve a competitive statement, we define a modified benchmark $\overline{\text{OPT}}$ that mirrors the same structural change as the online policy. We decompose the horizon into two stages. \emph{Stage 1} which considers the range $[1,\Bar{t}\!-\!1]$, where the benchmark $\overline{\text{OPT}}$ matches the online decisions by definition, and \emph{Stage 2} over $[\Bar{t},T]$, where the benchmark is the offline optimum for the reduced subproblem of discharging $k_{\rm rem}$. Thus, consider $R_{\rm stage 1}$ and $R_{\rm stage 2}$ as the online revenues in each of the previously defined stages, in particular, $R^{\rm bench}_{\rm stage 1}$ and $R^{\rm bench}_{\rm stage 2}$ are the rewards for the benchmark $\overline{\text{OPT}}$. By construction, $R^{\rm bench}_{\rm stage 1}\! = \!R_{\rm stage 1}$. At Stage 2, the online policy runs the threshold policy $G^{(k_{\rm rem})}$ on the remaining price sequence $p_{(\bar{t}:T)}$. The reward on this stage corresponds to:
\begin{equation}\label{eq:app_stage2_bound}
R_{\rm stage 2}\!= R\!\left(\!G^{(k_{\rm rem})}\!;\! p_{(\Bar{t}:T)}\!\right)\!\geq\!\frac{1}{\omega}
\text{OPT}^{(k_{\rm rem})}\!=\!\frac{1}{\omega}R^{\rm bench}_{\rm stage 2}.
\end{equation}

Finally, define the total revenues $R^{G}\!=\!R_{\rm stage 1}\!+\!R_{\rm stage 2}$ and $R^{\rm bench}\!=\!R^{\rm bench}_{\rm stage 1}\!+\!R^{\rm bench}_{\rm stage 2}$. Using \eqref{eq:app_stage2_bound}, $R^{G}$ satisfies:
\begin{align}
R^{G}
& \geq R_{\rm stage 1} + \tfrac{1}{\omega}R^{\rm bench}_{\rm stage 2} = \tfrac{1}{\omega} (\omega R_{\rm stage 1} + R^{\rm bench}_{\rm stage 2}), \nonumber\\
&\geq \tfrac{1}{\omega}(R^{\rm bench}_{\rm stage 1} + R^{\rm bench}_{\rm stage 2})
= \tfrac{1}{\omega}\,R^{\rm bench}.
\label{eq:app_two_stage_final}
\end{align}
Reducing $k$ to $k_{\rm rem}$ preserves the competitive guarantee as long as the benchmark is also forced to share the same two-stage structure and pre-change actions. Then, for any price sequence $p_{(1:T)}$ and two-stage structure, $R^{G} \geq \frac{1}{\omega}\,\overline{\text{OPT}}$.

\subsection{Feasibility-aware thresholds and pruned action sets}\label{app:feasible_pruned}
The SoC constraints restrict the number of discharge (and charge) actions that can be executed from a given state $(t,e_t)$. In particular, the maximum number of feasible discharge actions from $(t,e_t)$ is restricted as $K_{\rm max}^{\rm dis}(t,e_t)\!\!=\!\!\left\lfloor \frac{e_t - E_{\min}}{P} \right\rfloor$. This motivates us to replace the predicted-price set used in \eqref{NewThreshold} and \eqref{eq:app_time_dependent_thresh} with a feasibility-aware version. We replace the predicted-price set as follows:
\begin{equation}\label{eq:app_replace_pred_set}
\{\hat{z}^{\rm min}_m(t)\}_{m=1}^{k_{\rm dis}-(i-1)}
\;\;\longrightarrow\;\;
\{\hat{z}^{\rm min,feas}_m(t,e_t)\}_{m=1}^{K_{\rm rem}(t,e_t)},
\end{equation}
where $K_{\rm rem}(t,e_t) = \min\{K_{\max}^{\rm dis}(t,e_t), k_{\rm dis}-s(t)\}$ is the number of discharges still feasible, with $s(t)$ the number of discharges already executed along the previously realized sequence up to $t$. $\hat{z}^{\rm min,feas}_m(t,e_t)$ is the $m$-th smallest predicted price restricted to time slots that can host at least one feasible discharge along some feasible path. Under this modification, the feasibility-aware discharge threshold becomes:
\begin{equation} \label{eq:app_thresh_feas}
\ell_{i}^{\rm dis}(t,e_t)\!= \! \frac{\omega}{K_{\rm rem}(t,\!e_t)}
\!\Bigg(\! \sum_{\hat{t}=1}^{t-1}\sum_{n=1}^{i-1} \!\ell_{n}^{\rm dis \star}(\hat{t})
+\!\!\!\!\!\!\!\!\! \sum_{m=1}^{K_{\rm rem}(t,e_t)}\!\!\!\!\!\!\!\! \hat{z}_{m}^{\rm min,feas}(t,e_t) \Bigg).
\end{equation}
Thus, we can define the feasibility-aware benchmark $\text{OPT}_{f}$ at state $(t,e_t)$ as follows:
\begin{equation}\label{eq:app_bench_feas}
\text{OPT}_{f}^i(t,\!e_t) \!=\!\!\Bigg( \sum_{\hat{t}=1}^{t-1}\sum_{n=1}^{i-1}\! \ell_{n}^{\rm dis \star}(\hat{t}) +\!\!\!\!\!\!\sum_{m=1}^{K_{\rm rem}(t,e_t)} \!\!\!\!\!\hat{z}_{m}^{\rm min,feas}(t,\!e_t)\!\!\Bigg).
\end{equation}
By construction, the threshold in \eqref{eq:app_thresh_feas} can be written compactly and enforced to guarantee the competitive bound as: 
\begin{equation}\label{eq:app_u_equals_scaled_V}
\ell_{i}^{\rm dis}(t,e_t) = \tfrac{\omega}{K_{\rm rem}(t,e_t)}\,\text{OPT}_{f}(t,e_t).
\end{equation}
This construction preserves the same scaling logic as in \eqref{eq:app_thresh_vopt_relation}, but now enforces feasibility at the state level.

\end{document}